\documentclass[useAMS,usenatbib,onecolumn]{mn2e}
\usepackage{epsfig}
\usepackage{amssymb}
\usepackage{ifthen}

\newcommand{\ud}{{\rmn{d}}}
\newcommand{\vnhat}{\hat{\bmath{n}}}

\newcommand{\cle}{{\mathcal{E}}}
\newcommand{\clb}{{\mathcal{B}}}
\newcommand{\vl}{{\bmath{l}}}
\newcommand{\vL}{{\bmath{L}}}
\newcommand{\vx}{{\bmath{x}}}

\newcommand{\half}{{\textstyle{\frac{1}{2}}}}
\newcommand{\mybeth}{\bar{\eth}}
\newcommand{\lbar}{\bar{l}}

\title[Covariance of quadratic CMB power spectrum estimates]
{Error analysis of quadratic power spectrum estimates for CMB
polarization: sampling covariance}
\author[A.~Challinor and G.~Chon]
{Anthony Challinor$^1$\thanks{E-mail: a.d.challinor@mrao.cam.ac.uk}
and Gayoung Chon$^{1,2}$
\\
$^1$Astrophysics Group, Cavendish Laboratory, Madingley Road,
Cambridge CB3 0HE, U.K. \\
$^2$Computational Research Division, Lawrence Berkeley National Laboratory,
Berkeley, CA 94720, U.S.A.
}

\pagerange{\pageref{firstpage}--\pageref{lastpage}}
\pubyear{2004}

\label{firstpage}

\begin{document}

\maketitle

\begin{abstract}
Quadratic methods with heuristic weighting (e.g.\ pseudo-$C_l$ or
correlation function methods) represent an efficient way to
estimate power spectra of the cosmic microwave background (CMB) anisotropies
and their polarization. We construct the sample covariance properties of such
estimators for CMB polarization, and develop semi-analytic techniques
to approximate the pseudo-$C_l$ sample covariance matrices at high Legendre
multipoles, taking account of the geometric effects of mode coupling and the
mixing between the electric ($E$) and magnetic ($B$) polarization that arise
for observations covering only part of the sky. The $E$-$B$ mixing ultimately
limits the applicability of heuristically-weighted quadratic methods to
searches for the gravitational-wave signal in the large-angle $B$-mode
polarization, even for methods that can recover (exactly) unbiased estimates
of the $B$-mode power. We show that for surveys covering one or two per cent
of the sky, the contribution of $E$-mode power to the covariance of the
recovered $B$-mode power spectrum typically limits the tensor-to-scalar
ratio that can be probed with such methods to $\sim 0.05$.
\end{abstract}


\section{Introduction}
\label{sec:intro}

With the recent results from the Degree Angular Scale
Interferometer (DASI; \citealt{kovac02,leitch04}), the Cosmic Anisotropy
Mapper (CAPMAP; \citealt{barkats04}) and the Cosmic Background
Imager (CBI; \citealt{readhead04}), and the imminent two-year results
from the \emph{Wilkinson Microwave Anisotropy Probe}
(WMAP; \citealt{kogut03}), the characterization of the polarization
of the cosmic microwave background (CMB) polarization is gathering
pace. The ultimate goal of such experiments is to constrain parameters
describing the cosmological model and its fluctuations, and hence test
models for the generation of these fluctuations, such as inflation.
In Gaussian theories, all of the cosmological information encoded in the
CMB anisotropies and their polarization is contained in their power
spectra. For this reason, estimating CMB power spectra is a useful and
efficient form of data compression in analysis pipelines~\citep*{bond00}.
However, estimating the spectra is not enough: it is also essential to have
an accurate covariance matrix, and preferably an accurate model
for the shape of the likelihood function
(see e.g.~\citealt{bond00,efstathiou04}).
Modelling the covariance matrix for a particularly efficient class of
polarization power spectra estimators -- heuristically-weighted
quadratic estimators, or pseudo-$C_l$s -- is the subject of this paper.

The polarization field can be decomposed uniquely on the full sky
into a gradient part, with electric ($E$) parity, and a curl part with
magnetic ($B$) parity~\citep*{kamionkowski97,seljak97}. The cosmological
importance of this
decomposition is that linear density perturbations cannot produce
magnetic polarization, and so, in standard scenarios and at linear order,
large-angle $B$-mode polarization is produced solely by gravitational waves.
Second-order effects, most notably weak gravitational lensing by large-scale
structure, can generate $B$ modes from primary $E$ modes~\citep{zaldarriaga98}.
The lens-induced $B$ modes have a blue spectrum, and should dominate any
primary signal from gravitational waves on sub-degree scales. On larger
scales, they may still dominate if the gravitational wave amplitude is
sufficiently low. 
Either way, the variance of the $B$ modes is expected to be much smaller than
that of the $E$ modes.
Given the different parity properties of the $E$ and $B$ fields, CMB
polarization in a statistically-isotropic and parity-respecting universe
is characterized by two power spectra, $C_l^E$ and $C_l^B$, and
also the cross-power $C_l^{TE}$ between $E$ and the temperature anisotropies.

A number of different methods have been developed to estimate these 
CMB polarization power spectra. The optimal method, brute-force maximum
likelihood (e.g.\ \citealt*{gorski94,bond98}),
and its quadratic approximation~\citep{tegmark02},
should produce the most accurate (minimum-variance) estimates. However,
these methods require the inversion of $N_{\rmn{pix}}\times N_{\rmn{pix}}$
matrices, where $N_{\rmn{pix}}$ is the number of pixels in the map, which
is prohibitively slow for the analysis of mega-pixel maps. Given these
limitations, a number of fast alternative methods have been
suggested~\citep{hansen02b,kogut03,chon04} building on earlier work on
the analysis of the temperature
anisotropies~\citep*{szapudi01,wandelt01b,hivon02,hansen02}. These fast
methods are closely related, and all involve compressing heuristically-weighted
maps of the Stokes parameters into a
set of pseudo-$C_l$s making use of spherical transforms --
a step that requires only $O(N_{\rmn{pix}}^{3/2})$ operations with
fast Fourier transform methods on iso-latitude pixelisations such as the
widely-used HEALPix~\citep{gorski04}. The methods
differ in the details of how they transform the pseudo-$C_l$s to power spectrum
estimates. In the absence of instrument noise, the mean of the pseudo-$C_l$s
is linear in the true power spectra, so, for observations over a large enough
part of the sky, unbiased power spectrum estimates can be obtained by
applying the inverse operation to the pseudo-$C_l$s~\citep{kogut03}.
If the observed part of the sky is sufficiently small, direct inversion is
not possible and some form of regularisation is required. \citet{hansen02b}
advocate using likelihood methods, while~\citet{chon04} make use of correlation
functions as an intermediate data product. A notable feature of the latter
approach is that the inversion is constructed so that 
electric and magnetic polarization power is not mixed in the mean.
Pseudo-$C_l$ methods have now been applied to a number of real data sets,
both for temperature~\citep{netterfield02,hinshaw03} and
polarization analyses~\citep{kogut03,ponthieu05}.

Despite the attention that pseudo-$C_l$ techniques have received, relatively
little work has been done on trying to \emph{understand} their error
properties.
This paper aims to fill that gap by providing the basis for a reliable,
semi-analytic error analysis technique for CMB polarization power spectrum
estimates obtained with pseudo-$C_l$ methods. This extends earlier
analytic work that discusses temperature anisotropies
only~\citep{hinshaw03,efstathiou04}, and the recent thesis work by
one of us~\citep{chon_thesis} where both temperature and polarization are
considered.
A significant complication that arises for polarized data is the mixing
of $E$ and $B$ modes for observations covering only part of the
sky~\citep*{lewis02,bunn03}. We develop approximations to the pseudo-$C_l$
sample covariances, taking this mixing into account at leading order for maps
with smoothly-apodized edges. Our approximations are accurate in the regime
where the mixing is perturbative, i.e.\ for $l \gg l_{\rmn{max}}$, where
$l_{\rmn{max}}$ is the effective band-limit of the function used to weight
the Stokes parameters. It is difficult to
construct good approximations for the covariance on large scales,
$l \ll l_{\rmn{max}}$. This is partly because of the increased importance of
$E$-$B$ mixing there, but also because the polarization power spectra do not go
smoothly to zero on large scales in the presence of reionization. (The
behaviour of the temperature-anisotropy power spectrum on large scales
produces a similar effect for the temperature pseudo-$C_l$ covariance.)
Following~\citet{efstathiou04},
the low-$l$ sector of the covariance matrices can be constructed
directly with exact methods, and used to overwrite the inaccurate
approximation. Being able to evaluate accurate covariances efficiently is
important in any application where multiple evaluations are
required. Examples include parameter estimation, where the covariance
generally depends on position in parameter space, and optimising the design
and analysis of surveys, where one might wish to assess a large number of
weighting schemes.

Throughout, we ignore the effects of instrument noise so we deal only with
the sample covariance of the $\tilde{C}_l$. This is to highlight
the geometric effects of heuristic weighting that are peculiar to
polarization analysis, particularly $E$-$B$ mixing. An obvious extension of
this work that will be required for application to real data is to include the
contribution from instrument noise. For uncorrelated noise this should be
straightforward, but in a more realistic set-up, one will probably have to
resort to Monte-Carlo simulations for the noise contribution.

As noted above, one of the future goals of CMB polarization observations
is to detect (or constrain) the stochastic gravitational wave background
via its signature in large-angle $B$-mode
polarization~\citep{seljak97,kamionkowski97}. Magnetic polarization
allows much smaller amplitude gravitational wave backgrounds to be detected
than with temperature or electric polarization alone, since, in the
latter case, the gravitational-wave signal is swamped by the cosmic variance
of the signal from the dominant density perturbations. Methods to separate out
exactly pure magnetic modes from observations with incomplete sky coverage
exist at the map level~\citep{lewis02,bunn03,lewis03}. Such methods have the
desirable property that quadratic power spectrum estimates formed from
the pure-$B$ modes have no cosmic variance if the $B$-mode power is zero.
Were it not for gravitational lensing,
separating $B$ modes at the level of the map would thus allow the detection of
arbitrarily low amplitude gravitational wave backgrounds in the absence
of instrument noise. In practice, imperfections in removing the lensing
signal limit the amplitude to tensor-to-scalar ratios\footnote{Our
definition of the tensor-to-scalar ratio follows~\citet{martin00}, i.e.\
it is the ratio of the primordial gravitational wave and
curvature power spectra.} $r \sim 3\times 10^{-5}$
(\citealt*{kesden02,knox02}; see~\citealt{seljak03}
for refinements from more optimal lensing-reconstruction methods).
Heuristically-weighted pseudo-$C_l$ methods do not share the property
that the cosmic variance of the estimated $C_l^B$ is immune to $E$-mode
power, even if we take care to construct unbiased estimates so that e.g.\
the mean of our $C_l^B$ estimator decouples from $C_l^E$~\citep{chon04}.
We investigate this issue further here, exploring the limits of
detectability for $r$ that arise solely from cosmic variance of the $E$-mode
polarization leaking into the variance of quadratic estimates of
$C_l^B$.

This paper is organised as follows. Section~\ref{sec:properties}
reviews the properties of the polarization pseudo-$C_l$s and
establishes our notation. Following a brief discussion of the exact
pseudo-$C_l$ covariance matrices in Section~\ref{sec:exact-cov}, we
develop analytic approximations to them that can be rapidly computed
in Section~\ref{sec:approx-cov}. We begin with a toy model, based
on a Gaussian weight function applied to a small survey area in the flat-sky
limit. (Some of the more involved manipulations are relegated to
Appendix~\ref{app:flat}.)
This instructive example illustrates the key issues that are specific
to the analysis of polarization, i.e.\ the geometric mixing of $E$ and $B$
modes. We then move on to develop approximations for arbitrary (smooth)
weight functions, and test these against the exact covariances in
cases where we can easily compute the latter (in particular, for
azimuthally-symmetric weight functions). Simple rules-of-thumb for band-power
variances are also derived in Section~\ref{subsec:rot} as limiting cases of the
more general expressions;
some of the manipulations of the $3j$ symbols required for these derivations
are relegated to Appendix~\ref{app:useful_sums}.
Finally, in Section~\ref{sec:implications}
we consider the implications of our results for the detection of gravitational
waves with $B$-mode polarization estimated with pseudo-$C_l$ methods.

Throughout, we illustrate our results with the best-fitting power-law
concordance $\Lambda$CDM model to the \emph{Wilkinson Microwave
Anisotropy Probe} (\emph{WMAP}) and 2dF Galaxy Redshift Survey data
(\citealt{spergel03}; Table 7), but with the tensor-to-scalar ratio
$r=0.15$ corresponding to $\phi^2$ inflation. We include the effects of
gravitational lensing in the $B$-mode power spectrum, but ignore effects
due to its non-Gaussianity. 
Theoretical power spectra, pseudo-$C_l$s and
estimated power spectra in all figures include the smoothing effect of a
Gaussian beam of ten-arcmin FWHM.

\section{Properties of polarization pseudo-{$C_l$}s}
\label{sec:properties}

We define Stokes parameters along the line of sight $\vnhat$ using
the basis vectors of a spherical polar coordinate system. We take the
local $x$ axis to be $\hat{\bmath{\theta}}$ and the $y$ axis
$-\hat{\bmath{\phi}}$; these form a right-handed triad with the radiation
propagation direction $-\vnhat$ which defines the local $z$ axis.
The complex polarization $P \equiv Q + i U$ is then spin -2~\citep{newman66}
and can properly be expanded in terms of spin-2 spherical harmonics
(see e.g.~\citealt{lewis02} for a review):
\begin{eqnarray}
(Q\pm i U)(\vnhat) = \sum_{lm} (E_{lm} \mp i B_{lm}) {}_{\mp 2} Y_{lm}(\vnhat).
\label{eq:1}
\end{eqnarray}
Reality of the Stokes parameters ensures that $E_{lm}^\ast = (-1)^m E_{l-m}$
and similarly for $B_{lm}$, while under parity transformations
$E_{lm} \rightarrow (-1)^l E_{lm}$ (electric parity) but
$B_{lm} \rightarrow (-1)^{l+1} B_{lm}$ (magnetic parity). The polarization
power spectra are defined in statistically-isotropic and parity-respecting
models by
\begin{equation}
\langle E_{lm} E^*_{l'm'} \rangle = C_l^E \delta_{ll'} \delta_{mm'},
\quad \langle B_{lm} B^*_{l'm'} \rangle = C_l^B \delta_{ll'} \delta_{mm'}.
\label{eq:2}
\end{equation}
The $E$-$B$ power vanishes (and $B$ is uncorrelated with the temperature
anisotropies) by parity.

In pseudo-$C_l$ methods, one adopts a heuristic weighting of the complex
polarization with a scalar function $w(\vnhat)$. The weight is necessarily
zero in those regions that are not
observed, but typically it is desirable to excise a larger region to remove
further regions of high foreground emission. One can attempt to choose
$w(\vnhat)$ to improve the accuracy of subsequent power spectrum estimates.
For example, on small scales, where the instrument noise dominates the signal,
the optimal weighting is essentially inverse-noise-variance
weighting~\citep{efstathiou04}, while on large scales uniform weighting
is preferable. In this paper we shall assume that the weight is chosen
to be real and to go smoothly to zero as excised regions are approached.
The former restriction ensures that the weighting does not alter the
polarization direction, while the latter makes $w(\vnhat)$
approximately band-limited to some $l_{\rmn{max}}$.
Such edge apodization allows us to deal with the effects of the
resulting $E$-$B$ mixing in the pseudo-$C_l$ covariance matrices in
a perturbative manner; see Section~\ref{sec:approx-cov}. The problem of
$E$-$B$ mixing without edge apodization, i.e.\ taking $w(\vnhat)$ everywhere
one or zero, is discussed extensively by~\citet{lewis02} and~\citet{lewis03}
in the context of isolating $B$ modes at the map level.

Following~\citet{hansen02b}, we define pseudo-multipoles $\tilde{E}_{lm}$ and
$\tilde{B}_{lm}$ to be the $E$ and $B$-mode multipoles of the weighted
polarization map, so that
\begin{equation}
\tilde{E}_{lm} \pm i \tilde{B}_{lm}
= \sum_{(lm)'} {}_{\pm2}I_{(lm)(lm)'} (E_{(lm)'}\pm iB_{(lm)'}),
\label{eq:5-25}
\end{equation}
where ${}_{\pm 2}I_{(lm)(lm)'}$ are the Hermitian coupling matrices
\begin{equation}
{}_{\pm 2} I_{(lm)(lm)'} = \int \ud\vnhat \, w(\vnhat)
{}_{\pm 2} Y_{(lm)'}(\vnhat){}_{\pm 2} Y_{lm}^* (\vnhat).
\label{eq:4}
\end{equation}
Using ${}_s Y_{lm}^* = (-1)^{s+m} {}_{-s}Y_{l-m}$, and the reality of
$w(\vnhat)$, we have ${}_{\pm 2} I_{(lm)(lm)'}^* = (-1)^{m+m'}
{}_{\mp 2} I_{(l-m)(l-m)'}$ which ensures that $\tilde{E}_{lm}^* =
(-1)^m \tilde{E}_{l-m}$ and a similar result for $\tilde{B}_{lm}$.
If we expand the weight function in terms of its spherical multipoles,
$w(\vnhat) = \sum_{lm} w_{lm} Y_{lm}(\vnhat)$, the coupling matrices
can be evaluated in terms of Wigner-$3j$ symbols:
\begin{equation}
{}_{\pm 2} I_{(lm)(lm)'} = \sum_{LM} (-1)^m w_{LM}
\sqrt{\frac{(2L+1)(2l+1)(2l'+1)}{4\pi}} 
\left( \begin{array}{ccc} l & l' & L \\ m & -m' & -M \end{array} \right) 
\left( \begin{array}{ccc} l & l' & L \\ \mp 2&\pm 2 &0 \end{array} \right). 
\label{eq:5-26}
\end{equation}
From equation~(\ref{eq:5-25}) we can separate the $E$ and $B$
pseudo-multipoles as 
\begin{eqnarray}
\tilde{E}_{lm} &=& \sum_{(lm)'} 
\left( {}_+ I_{lm(lm)'}E_{(lm)'} + i{}_{-}I_{(lm)(lm)'}B_{(lm)'} \right),
\label{eq:5-27}\\
\tilde{B}_{lm} &=& \sum_{(lm)'}
\left( {}_+ I_{(lm)(lm)'}B_{(lm)'} - i{}_{-}I_{(lm)(lm)'}E_{(lm)'} \right),
\label{eq:5-28}
\end{eqnarray}
where we define the Hermitian ${}_\pm I_{(lm)(lm)'}$ to be
\begin{equation}
{}_+ I_{(lm)(lm)'} \equiv \frac{1}{2}({}_{+2} I_{(lm)(lm)'}
+ {}_{-2}I_{(lm)(lm)'}),
\qquad {}_{-}I_{(lm)(lm)'} \equiv \frac{1}{2}({}_{+2}I_{(lm)(lm)'} -
{}_{-2}I_{lm(lm)'}).
\label{eq:5-29}
\end{equation}
Given the $\tilde{E}_{lm}$ and $\tilde{B}_{lm}$, we form their pseudo-$C_l$s
according to
\begin{equation}
\tilde{C}_l^E \equiv \frac{1}{2l+1}\sum_{m} |\tilde{E}_{lm}|^2, \quad
\tilde{C}_l^B \equiv \frac{1}{2l+1}\sum_{m} |\tilde{B}_{lm}|^2.
\label{eq:9}
\end{equation}
We could also form $\tilde{C}_l^{EB} \equiv \sum_m \tilde{E}_{lm}
\tilde{B}_{lm}^*/(2l+1)$, but, as we show shortly, this vanishes in the
mean in the absence of parity violations. However, it may provide a
useful test on parity-violating physics or unaccounted-for systematics,
and so future data should, of course, be validated with $\tilde{C}_l^{EB}$.

In the absence of instrument noise, the mean pseudo-$C_l$s (i.e.\
averaged over CMB realisations) are related to the true power spectra
by convolutions. We find~\citep{hansen02b}
\begin{equation}
\langle \tilde{C}_l^E \rangle = \sum_{l'} (P_{ll'} C_{l'}^E +
M_{ll'} C_{l'}^B),
\quad
\langle \tilde{C}_l^B \rangle = \sum_{l'} (M_{ll'} C_{l'}^E +
P_{ll'} C_{l'}^B),
\label{eq:10}
\end{equation}
where
\begin{eqnarray}
P_{ll'} &\equiv& \frac{1}{2l+1} \sum_{mm'} |{}_+ I_{(lm)(lm)'}|^2
        = \frac{2l'+1}{8\pi} \sum_L (2L+1) w_L [1+(-1)^K]
\left( \begin{array}{ccc} l & l' & L \\ -2 & 2 &0 \end{array} \right)^2 ,
\label{eq:11} \\
M_{ll'} &\equiv& \frac{1}{2l+1} \sum_{mm'} |{}_- I_{(lm)(lm)'}|^2
        = \frac{2l'+1}{8\pi} \sum_L (2L+1) w_L [1-(-1)^K]
\left( \begin{array}{ccc} l & l' & L \\ -2 & 2 &0 \end{array} \right)^2 .
\label{eq:12}
\end{eqnarray}
Here $K \equiv l+l'+L$ and we have introduced the power spectrum of the
weight function $w_l \equiv \sum_m |w_{lm}|^2 / (2l+1)$. A non-zero matrix
$M_{ll'}$ biases the pseudo-$C_l$s by mixing e.g.\ $E$-mode power into
$\tilde{C}_l^B$. For observations with uniform weight over the full sky,
$w(\vnhat)=\rmn{constant}$ and $M_{ll'}=0$ since then only $w_0$ is non-zero
and the $3j$ symbol forces $l=l'$ and hence $K$ to be even. The sum of
$P_{ll'}$ and $M_{ll'}$ is approximately equal to the matrix that relates
the temperature $C_l$s to the mean pseudo-$C_l$s~\citep{hivon02} at high
$l$. Their sum differs from the temperature case only by the presence of
non-zero azimuthal quantum numbers $\pm 2$ in the $3j$ symbol.
The mean of $\tilde{C}_l^{EB}$ evaluates to
\begin{equation}
\langle \tilde{C}_l^{EB} \rangle = \sum_{l'} (P_{ll'} - M_{ll'}) C_{l'}^{EB},
\label{eq:13}
\end{equation}
and so vanishes if parity is preserved in the mean ($C_l^{EB}=0$).

\begin{figure}
\begin{center}
\epsfig{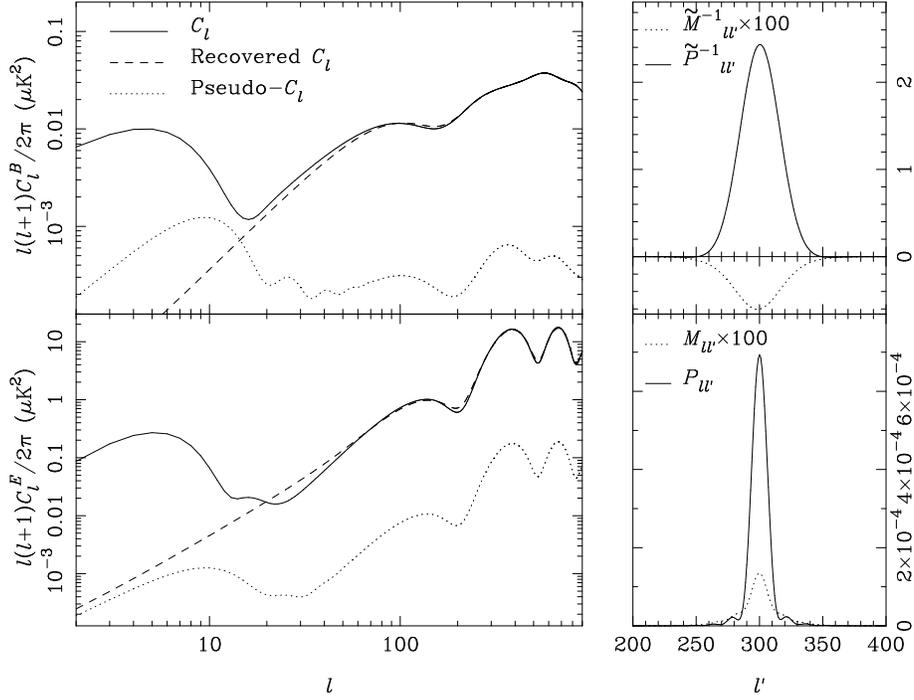}
\end{center}
\caption{Left: power spectra for $B$ (top) and $E$ (bottom; solid lines)
compared to the mean pseudo-$C_l$s (dotted lines) and the recovered power
spectra (dashed lines). Right: the bottom panel shows representative window
functions $P_{ll'}$ (solid lines) and $M_{ll'}$ (dotted lines) that
give the mean pseudo-$C_l$s on convolving with the true $C_l$s; the top panel
shows the pseudo-inverses $\tilde{P}^{-1}_{ll'}$ (solid lines) and
$\tilde{M}^{-1}_{ll'}$ (dotted lines) that when convolved with the
mean pseudo-$C_l$s remove the effect of $E$-$B$ mixing. Note that
$M_{ll'}$ and $\tilde{M}^{-1}_{ll'}$ have been multiplied by a factor of
100 for clarity. The weight function applied to the map is uniform inside
a circle of $10\degr$ radius, with cosine apodization out to $15\degr$.
To obtain the pseudo-inverses, a Gaussian apodization of $4\degr$ HWHM is
applied to the correlation functions.
\label{fig:power_and_window}
}
\end{figure}

An example of the mean pseudo-$C_l$s for observations over a circular
region of radius $15\degr$ is shown in Fig.~\ref{fig:power_and_window}.
The weight function is uniform inside a circle of radius $10\degr$, but
the remaining $5\degr$ is apodized as $\cos^2 [36(\theta-10\degr)]$ so that
$w(\vnhat)$ falls smoothly to zero at the edge of the observed region.
For this weight function the matrix $P_{ll'}$ is roughly two orders of
magnitude larger than $M_{ll'}$, and both are narrow compared to features
in $C_l^E$ and $C_l^B$ except for the reionization signal on the largest
scales. On small scales the mean pseudo-$C_l$s can be approximated by
removing $C_{l}^E$ and $C_{l}^B$ from the convolutions in
equation~(\ref{eq:10}) to give
\begin{equation}
\langle \tilde{C}_l^E \rangle \approx C_l^E \sum_{l'} P_{ll'} + C_l^B
\sum_{l'} M_{ll'} , \quad
\langle \tilde{C}_l^B \rangle \approx C_l^B \sum_{l'} P_{ll'} + C_l^E
\sum_{l'} M_{ll'}. 
\label{eq:5-36} 
\end{equation}
For $\tilde{C}_l^E$ the mixing term described by $M_{ll'}$ is essentially
negligible, and the mean of $\tilde{C}_l^E$ is approximately a scaled version
of the underlying power spectrum. The relevant scaling is discussed below.
For $\tilde{C}_l^B$ mixing is important, despite $P_{ll'}$ dominating over
$M_{ll'}$, since $C_l^E \gg C_l^B$. This additional contribution to $\langle
\tilde{C}_l^B \rangle$ from geometric mixing of $E$ and $B$ modes can be
clearly seen in Fig.~\ref{fig:power_and_window}, and, as expected, looks like
a scaled version of the $E$-mode power spectrum.

The normalisations $\sum_{l'} P_{ll'}$ and $\sum_{l'}M_{ll'}$ evaluate to
\begin{eqnarray}
\sum_{l'} P_{ll'} + M_{ll'} &=& \sum_L \frac{2L+1}{4\pi} w_L 
= w^{(2)} f_{\rmn{sky}},
\label{eq:5-35a}
\\
\sum_{l'} P_{ll'} - M_{ll'} &=& \sum_{L \leq l}
\frac{2L+1}{4\pi} w_L
\left( 1-4\frac{L(L+1)}{l(l+1)} + 3\frac{(L+2)!}{(L-2)!}
\frac{(l-2)!}{(l+2)!} \right),
\label{eq:5-35}
\end{eqnarray}
where $w^{(2)}f_{\rmn{sky}}$ is defined by $4\pi w^{(i)}
f_{\rmn{sky}} \equiv \int w^i(\vnhat) \, \ud \vnhat$,
and $f_{\rmn{sky}}$ is the fraction of sky
with non-zero weighting. Equation~(\ref{eq:5-35a}) follows from
the expansions of $P_{ll'}$ and $M_{ll'}$ in
equations~(\ref{eq:11}) and~(\ref{eq:12}), and the orthogonality of the
$3j$ symbols. It is the same normalisation as for the temperature
anisotropies. Equation~(\ref{eq:5-35}) can be established following the
method used to derive equation (66) in~\citet{chon04}; see also
Appendix~\ref{app:useful_sums}. Note that for $l$ much greater than the
assumed band-limit of $w(\vnhat)$, the normalisation $\sum_{l'} M_{ll'}$ is
suppressed by a factor $O(l_{\rmn{max}}^2/l^2)$ compared to $\sum_{l'}
P_{ll'}$. This is consistent with the expectation that $E$-$B$ mixing
is suppressed at high $l$~\citep{lewis02,bunn03}. We can gain further insight
into this result by noting that [for smooth $w(\vnhat)$]
\begin{eqnarray}
\sum_{L \leq l} (2L+1) L(L+1) w_L &=& \int |\nabla
w_{\rmn{s},l}|^2 \, \ud\vnhat , \label{eq:14a} \\
\sum_{L \leq l} (2L+1) \frac{(L+2)!}{(L-2)!}
w_L &=& \int w_{\rmn{s},l} \nabla^2(\nabla^2+2) w_{\rmn{s},l} \, \ud\vnhat ,
\label{eq:14b}
\end{eqnarray}
where $w_{\rmn{s},l}(\vnhat) \equiv \sum_{L\leq l, M}w_{LM} Y_{LM}(\vnhat)$
is $w(\vnhat)$ smoothed with a top-hat ($l$-space) filter. For band-limited
weight functions, at $l \gg l_{\rmn{max}}$ the smoothing has little effect
and so to leading order
\begin{equation}
\sum_{l'} M_{ll'} \sim \frac{1}{2\pi} \int \frac{|\nabla w|^2}{l(l+1)}
\, \ud \vnhat\quad (l \gg l_{\rmn{max}}) .
\label{eq:15}
\end{equation}
At a given scale $l$, the relative importance of mixing of $E$- and $B$-mode
power in the pseudo-$C_l$s is thus controlled by $1/l^2$ times the ratio of the
average squared gradient of the weight function to the average square
weight function.

\subsection{Inversion of the pseudo-$C_{l}$s}
\label{sec:inverse}

For the purpose of parameter estimation, it is not strictly necessary to
attempt to invert the pseudo-$C_l$s to (unbiased) power spectrum estimates.
Forming the pseudo-$C_l$s can be regarded as a mildly-lossy form of data
compression of the maps of Stokes parameters, and inferences about
cosmological models can be made directly from them if their covariance -- or,
better, their full sampling distribution -- can be
calculated~\citep{wandelt01b}. The calculation of the signal contribution
to these covariances is the main topic of this paper.
However, for presentation purposes it is useful to be able to plot
quantities that are simply related to the underlying power spectra. At the
least this requires some approximate deconvolution of the geometric
effects of the weight function from the pseudo-$C_l$s.

For observations covering enough of the sky\footnote{%
The criterion for invertibility is equivalent to being able
to estimate the correlation functions for all angular
separations~\citep{chon04}.
A sufficient condition is thus that the observed region
has pixels separated by all angles between 0 and $180\degr$.},
the matrices $P_{ll'}$ and $M_{ll'}$ will be invertible. In this case we
easily obtain unbiased estimates of $C_l^E$ and $C_l^B$ from
\begin{equation}
\hat{C}_l^E + \hat{C}_l^B = \sum_{l'} (P+M)^{-1}_{ll'}(\tilde{C}_l^E +
\tilde{C}_l^B), \quad
\hat{C}_l^E - \hat{C}_l^B = \sum_{l'} (P-M)^{-1}_{ll'}(\tilde{C}_l^E -
\tilde{C}_l^B).
\label{eq:16}
\end{equation}
Alternatively, we can form unbiased estimates of the correlation functions
of the Stokes parameters from the polarization pseudo-$C_l$s (or directly
from the maps), and then invert these with the appropriate Legendre
transforms~\citep{chon04}. For observations over smaller parts of the sky,
$P_{ll'}$ and $M_{ll'}$ will not be invertible. In this case we do not
have the spectral resolution to obtain unbiased estimates of the $C_l$s
at every $l$. Various approaches have been suggested to regularise the
inversion including the use of band-powers~\citep{hivon02,hansen02}, or
apodized Legendre transforms of the (incomplete) correlation
functions~\citep{szapudi01,chon04}. The latter method provides estimates
of the $C_l$s convolved with a window function that depends only on the choice
of apodizing function that is applied to the correlation function.
\citet{chon04}, building on earlier work by~\citet*{crittenden02}, showed
how to generalise the correlation function method to ensure that the
estimated polarization power spectra did not mix $E$ and $B$ power in the
mean. This is particularly useful for visualisation of the $B$-mode
power spectrum which can easily be swamped by the dominant $E$ modes.
We do not reproduce the method of~\citet{chon04} here, but note only that
it provides a convenient way of constructing pseudo-inverses
$\tilde{P}^{-1}_{ll'}$ and $\tilde{M}^{-1}_{ll'}$ to $P_{ll'}$ and
$M_{ll'}$ respectively with the properties that
\begin{equation}
\sum_{l'} (\tilde{P}^{-1}_{ll'} P_{l'l''} +\tilde{M}^{-1}_{ll'} M_{l'l''})
= {}_{-2}\bar{K}_{ll''} , \quad
\sum_{l'} (\tilde{P}^{-1}_{ll'} M_{l'l''} +\tilde{M}^{-1}_{ll'} P_{l'l''})
= 0. 
\label{eq:17}
\end{equation}
These relations ensure that our estimates of the power spectra,
\begin{equation}
\hat{C}_l^E = \sum_{l'} (\tilde{P}^{-1}_{ll'} \tilde{C}_{l'}^E +
\tilde{M}^{-1}_{ll'} \tilde{C}_{l'}^B), \quad
\hat{C}_l^B = \sum_{l'} (\tilde{P}^{-1}_{ll'} \tilde{C}_{l'}^B +
\tilde{M}^{-1}_{ll'} \tilde{C}_{l'}^E),
\label{eq:estimates}
\end{equation}
satisfy $\langle \hat{C}_l^E \rangle = \sum_{l'} {}_{-2}\bar{K}_{ll'} C_{l'}^E$
and the equivalent relation for $C_l^B$. The normalised window function
${}_{-2}\bar{K}_{ll'}={}_{-2} K_{ll'} / \sum_L {}_{-2} K_{lL}$, where
${}_{-2}K_{ll'}$ is given by~\citep{chon04}
\begin{equation}
{}_{-2}K_{ll'} \equiv \frac{2l'+1}{2} \int f(\beta) d^l_{2\,-2}(\beta)
d^{l'}_{2\,-2}(\beta) \, \ud \cos\beta.
\end{equation}
Here, $d^l_{mn}(\beta)$ are the reduced Wigner rotation matrices and
$f(\beta)$ is the apodizing function that is applied to the correlation
functions. An example of the pseudo-inverses obtained via this route
is shown in Fig.~\ref{fig:power_and_window}, along with the mean of the
recovered $C_l$s. The correlation functions have been apodized with
a Gaussian with HWHM of $4\degr$. The apodizing function was deliberately
chosen to be narrow to ensure that the resulting pseudo-inverses
$\tilde{P}^{-1}_{ll'}$ and $\tilde{M}^{-1}_{ll'}$ are well localised. Given
that the pseudo-inverses are significantly broader than $P_{ll'}$ and
$M_{ll'}$, the window function ${}_{-2}K_{ll'}$ (not shown in the figure)
inherits the width of the pseudo-inverses. For this reason the spectral
resolution of the recovered $C_l$s is rather poor compared to what should
ultimately be achievable for a $15\degr$-radius survey. The effect of this
is that the recovered $C_l$s accurately follow the true
$C_l$s only for $l \ga 50$.

\section{Exact covariance properties}
\label{sec:exact-cov}

The covariance of any quadratic power spectrum estimates that derive
from the pseudo-$C_l$s by linear transformations follow simply from
the covariance of the latter. For this reason, in this paper we
concentrate on the pseudo-$C_l$ covariance. Furthermore, we focus on the
sample covariance (i.e.\ we neglect instrument noise) to highlight those
aspects specific to CMB polarization that arise from mode-mixing on the
incomplete sky. Including instrument noise is obviously important
for application to near-future surveys since it will dominate sample
variance. Extending the analytic results derived in this paper to include
simple white noise should be straightforward (see~\citealt{efstathiou04}
for the case of the temperature anisotropies). However, for more realistic
correlated noise, Monte-Carlo simulations will probably be required
to account properly for the noise contribution to the covariances.
We shall assume throughout that the CMB signal is Gaussian. If the initial
fluctuations were very-nearly Gaussian, as expected in simple single-field
inflation models (e.g.~\citealt{bartolo04}), non-linear effects will
introduce only a small level of non-Gaussianity into $E$ modes on
the scales of interest for primary CMB science (e.g.~\citealt*{mollerach04}).
However,
if the primordial tensor-to-scalar ratio $r \la 0.01$, then the dominant
power in $B$ modes is expected to come from gravitational-lensing conversion of
Gaussian $E$ modes into non-Gaussian $B$ modes. In this limit, our assumption
of Gaussianity will be violated and the expressions derived in this paper
will require some modification to account for the non-zero connected four-point
function. While a full treatment of the pseudo-$C_l$ covariance in the
presence of lens-induced $B$ modes is beyond the scope of the current paper,
we note that the lensing contribution to $\rmn{cov}(\tilde{C}_l^B,
\tilde{C}_{l'}^B)$ has been calculated recently for uniform weighting
in the flat-sky limit~\citep*{smith04}.

In this section we calculate the exact sample covariance
of $\tilde{C}_l^E$ and $\tilde{C}_l^B$ (see also~\citealt{hansen02b}
for a similar calculation) and present the results of
numerical computations for two types of survey: a full-sky survey with
a simple Galactic cut, and a $15\degr$-radius circular survey of the
type that is optimal for gravitational wave searches in the presence of
instrument noise at the level achievable with current
technology~\citep{jaffe00,lewis02}.

For Gaussian CMB fields, the pseudo-multipoles are also Gaussian, so
the covariance of their power spectra are given by e.g.\
\begin{equation}
\rmn{cov}(\tilde{C}_l^E,\tilde{C}_{l'}^E) = \frac{2}{(2l+1)(2l'+1)}
\sum_{mm'} | \langle \tilde{E}_{lm} \tilde{E}_{(lm)'}^* \rangle |^2 .
\end{equation}
If we expand the pseudo-multipoles in terms of the true multipoles using
equation~(\ref{eq:5-25}), we find
\begin{eqnarray}
{\rmn{cov}} (\tilde{C}_l^E,\tilde{C}_{l'}^E )
&=& \frac{2}{(2l+1)(2l'+1)} \sum_{mm'} \left|\sum_{LM}
{}_{+}I_{(lm)(LM)}{}_{+}I^*_{(lm)'(LM)}C_L^E + {}_{-}I_{(lm)(LM)}
{}_{-}I^*_{(lm)'(LM)}C_L^B \right|^2,
\label{eq:5-37}\\
{\rmn{cov}} (\tilde{C}_l^B,\tilde{C}_{l'}^B ) 
&=& \frac{2}{(2l+1)(2l'+1)} \sum_{mm'} \left|\sum_{LM}
{}_{+}I_{(lm)(LM)} {}_{+}I^*_{(lm)'(LM)}C_L^B + {}_{-}I_{(lm)(LM)}
{}_{-}I^*_{(lm)'(LM)}C_L^E \right|^2, \label{eq:5-38} \\
{\rmn{cov}} (\tilde{C}_l^E,\tilde{C}_{l'}^B ) 
&=& \frac{2}{(2l+1)(2l'+1)} \sum_{mm'} \left|\sum_{LM}
{}_{+}I_{(lm)(LM)} {}_{-}I^*_{(lm)'(LM)}C_L^E + {}_{-}I_{(lm)(LM)}
{}_{+}I^*_{(lm)'(LM)}C_L^B
 \right|^2.
\label{eq:5-39}
\end{eqnarray}
Evaluating these expressions numerically is
straightforward in principle, but is computationally expensive, with
an operation count of $O(l^6)$ if multipoles up to $l$ are retained.
This is prohibitive for $l$ above a few hundred, and so, in general,
there is a need for faster approximate calculations. One such approximation
is to compute the covariances from Monte-Carlo simulations; this has the
benefit of allowing one easily to include a number of real-world effects that
are more difficult to include in the analytic calculation. An alternative,
which is developed here in Section~\ref{sec:approx-cov}, is to use
analytic approximations (that can be calibrated with simulations if
required). As we show later, it is difficult to find good analytic
approximations at low $l$ because of the expected sharp rise in the
polarization $C_l$s there, and the increased importance of $E$-$B$ mixing.
A similar problem arises for the temperature anisotropies
because the $C_l$s do not go smoothly to zero as $l \rightarrow 2$.
\citet{efstathiou04} suggests replacing the block of the
approximate covariance matrices with small $l$ and $l'$ with the exact
expression, since the latter can be
calculated quickly at small $l$ and $l'$. For diagonally-dominant matrices,
this procedure returns an accurate covariance matrix; a similar procedure can
be implemented for polarization using
equations~(\ref{eq:5-37})--(\ref{eq:5-39}) and the approximations developed
in Section~\ref{sec:approx-cov}.

\subsection{Examples}
\label{subsec:ex1}

As our first application of equations~(\ref{eq:5-37})--(\ref{eq:5-39}),
we consider a full-sky survey from which we remove
a $\pm 20\degr$ band around the Galactic plane. We adopt a north-south
symmetric weighting which is uniform
(which we expect to be nearly optimal in the noise-free limit)
except for the edge of the retained survey which we apodize with the
square of a cosine.
In the northern hemisphere
\begin{equation}
w(\theta,\phi) = \left\{ \begin{array}{ll} 1 \quad & \mbox{for
$\theta < 65\degr$} \\
\cos^2[36(70\degr-\theta)] & \mbox{for $65\degr \le \theta \le 70\degr$.}
\end{array} \right.
\label{eq:exact_cov1}
\end{equation}
Since $w(\vnhat)$ is azimuthally-symmetric, the matrices $\pm I_{(lm)(lm')}$
are block diagonal and the covariances can be evaluated exactly
in $O(l^4)$ operations.

\begin{figure}
\begin{center}
\epsfig{figure=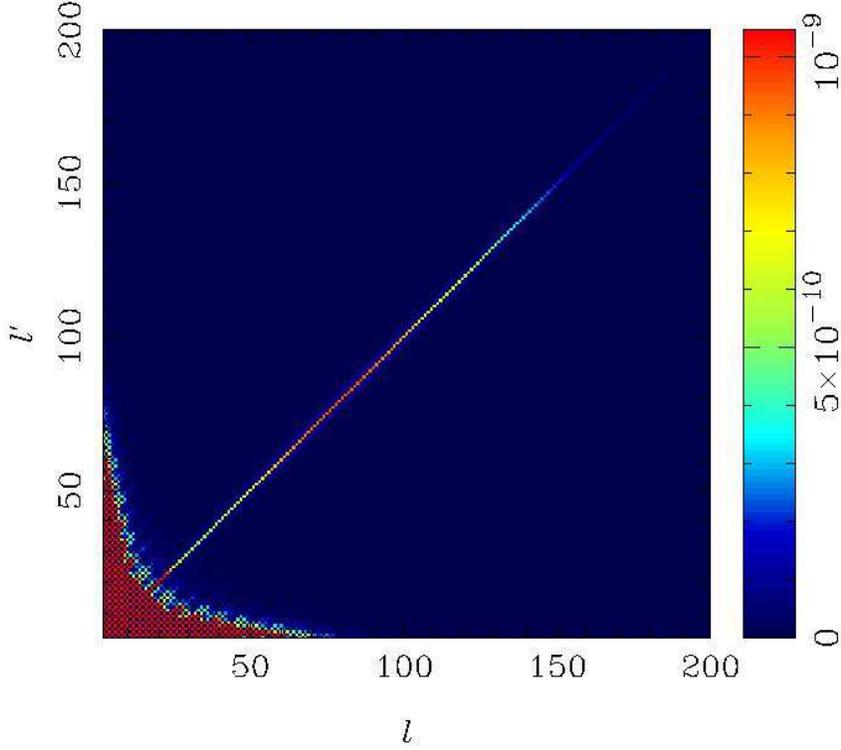,angle=0,width=12cm}
\end{center}
\caption{The covariance matrix
$\rmn{cov}(\tilde{C}_l^E, \tilde{C}_{l'}^E)$ for a
Galactic cut of $\pm 20\degr$ apodized with the square of a cosine over a
width of $5\degr$. The oscillatory structure at low $l$ (which
saturates the colour scale) is due to the reionization
feature in $C_l^E$ and residual Fourier ringing in ${}_+I_{(lm)(lm)'}$.
In this plot, as with all plots of covariance matrices in this paper,
the units are $\mu\rmn{K}^2$.
\label{fig:exactEE_galcosine}
}
\end{figure}

\begin{figure}
\epsfig{figure=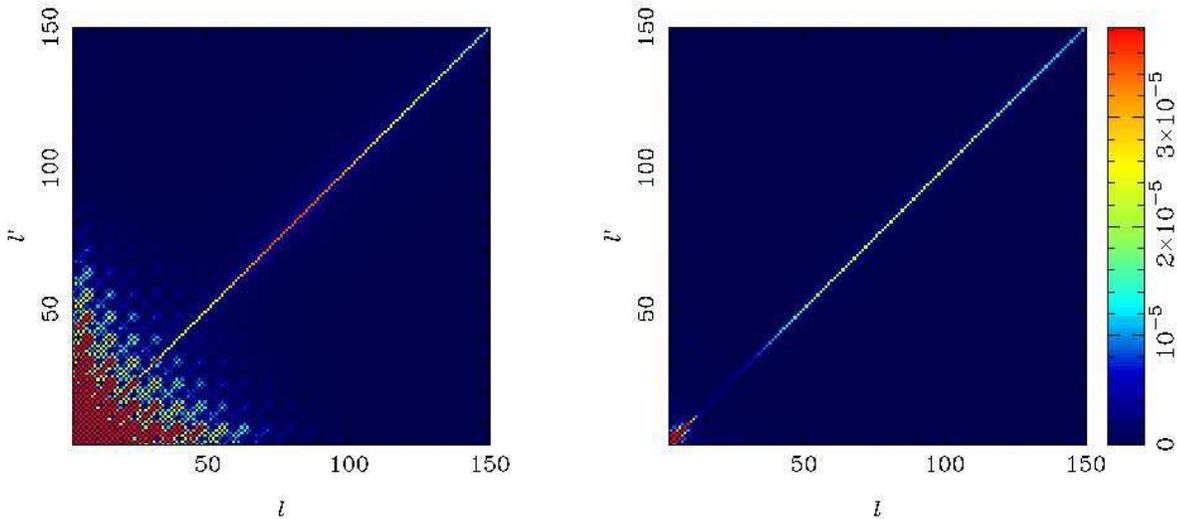,angle=0,width=16cm}
\caption{The covariance matrix
$l(l+1)l'(l'+1)\rmn{cov}(\tilde{C}_l^B, \tilde{C}_{l'}^B)$ for the
same Galactic cut as in Fig.~\ref{fig:exactEE_galcosine}.
The left panel is the exact covariance matrix, while the right
panel has $C_l^E=0$ to show the importance of $E$-$B$ mixing for this
covariance matrix. We have multiplied by $l(l+1)l'(l'+1)$ to make the
diagonal elements more nearly uniform.
The oscillatory structure in the matrix on the left at
low $l$ is due to the interaction of the reionization feature in $C_l^E$
with residual Fourier ringing in the coupling matrix ${}_-I_{(lm)(lm)'}$.
\label{fig:exactBB_galcosine}
}
\end{figure}

\begin{figure}
\begin{center}
\epsfig{figure=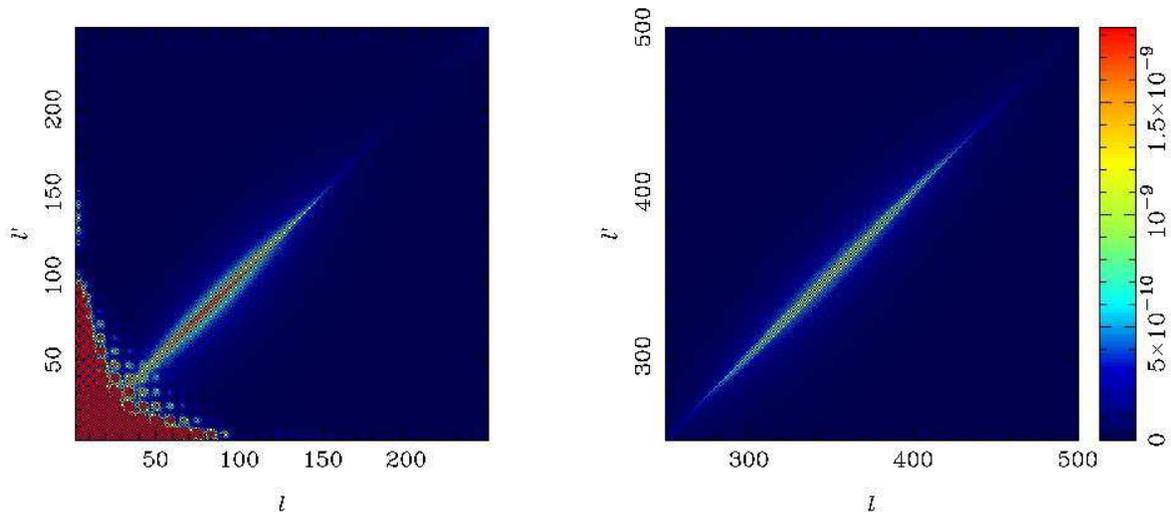,angle=0,width=16cm}
\end{center}
\caption{Blocks of the covariance matrix
$ll'\rmn{cov}(\tilde{C}_l^E, \tilde{C}_{l'}^B)$ for the
same Galactic cut as in Fig.~\ref{fig:exactEE_galcosine}.
We have multiplied by $l l'$ to make the elements near the diagonal more
nearly uniform.
\label{fig:exactEB_galcosine}
}
\end{figure}


Figures~\ref{fig:exactEE_galcosine}--\ref{fig:exactEB_galcosine} show the
exact covariance matrices $\rmn{cov}(\tilde{C}_l^E,\tilde{C}_{l'}^E)$,
$l(l+1)l'(l'+1)\rmn{cov}(\tilde{C}_l^B,\tilde{C}_{l'}^B)$ and
$ll'\rmn{cov}(\tilde{C}_l^E,\tilde{C}_{l'}^B)$ respectively. The prefactors
have been chosen to make the elements near the diagonal more nearly uniform.
Given that the weight function is symmetric under parity,
${}_+ I_{(lm)(lm)'} = (-1)^{l+l'} {}_+ I_{(lm)(lm)'}$ and
${}_- I_{(lm)(lm)'} = -(-1)^{l+l'} {}_- I_{(lm)(lm)'}$, so that
$\rmn{cov}(\tilde{C}_l^E,\tilde{C}_{l'}^E)$ and $\rmn{cov}(\tilde{C}_l^B,
\tilde{C}_{l'}^B)$ vanish if $l+l'$ is odd, while
$\rmn{cov}(\tilde{C}_l^E,\tilde{C}_{l'}^B)$ vanishes if $l+l'$ is even
(and hence on the diagonal).
In each of the three covariance matrices, the intense feature for
$l$ and $l' \la 50$ (which
saturates the colour scale) arises predominantly from the additional power in
$C_l^E$ on large scales due to reionization. (It vanishes if we
substitute for power spectra with no reionization.)

In the case of
$\rmn{cov}(\tilde{C}_l^E,\tilde{C}_{l'}^E)$, the first term inside the
summation over $L$ and $M$ in equation~(\ref{eq:5-37}) is always dominant.
The covariance is thus very similar to that
for the temperature anisotropies with the same weight function. The matrix
is diagonally dominant, except on large scales, with width inversely related
to the angular dimension of the retained region.
On large scales the covariance matrix is dominated by the additional
polarized power generated by reionization. Note that this power propagates
to somewhat smaller scales in the pseudo-$C_l$ covariance due to residual
Fourier ringing in the matrix ${}_+ I_{(lm)(lm)'}$. (For the weight function
adopted here, the width of the
main peak of ${}_\pm I_{(lm)(lm)'}$ is given by the inverse of the survey size,
but the effective band-limit is larger being set by the inverse size of the
apodized region.)

To emphasise the importance of $E$-$B$ mixing
in the structure of the covariance matrix for $\tilde{C}_l^B$, in
Fig.~\ref{fig:exactBB_galcosine} we show both the exact
$\rmn{cov}(\tilde{C}_l^B,\tilde{C}_{l'}^B)$ and the matrix
we obtain by setting $C_l^E=0$. The latter is very inaccurate on the
diagonal at $l \la 120 $, where $E$-$B$ mixing is most important, and
particularly so at the largest scales where it further fails to pick up the
additional power in $C_l^E$ from reionization. Setting $C_l^E$ to zero
also artificially suppresses the correlations: there is significantly more
structure off the diagonal in the left-hand plot in
Fig.~\ref{fig:exactBB_galcosine} than on the right. This reflects the
fact that ${}_-I_{(lm)(lm)'}$ is intrinsically broader than
${}_+ I_{(lm)(lm')}$ (see Fig.~\ref{fig:power_and_window}).

The covariance matrix $\rmn{cov}(\tilde{C}_l^E,\tilde{C}_{l'}^B)$, shown in
Fig.~\ref{fig:exactEB_galcosine}, is only non-zero because of $E$-$B$ mixing.
It is correspondingly broader than $\rmn{cov}(\tilde{C}_l^E,\tilde{C}_{l'}^E)$.
The dominant contribution is from $E$-mode power, so the covariance matrix
peaks near the diagonal at the positions of the acoustic peaks in $C_l^E$.

\begin{figure}
\epsfig{figure=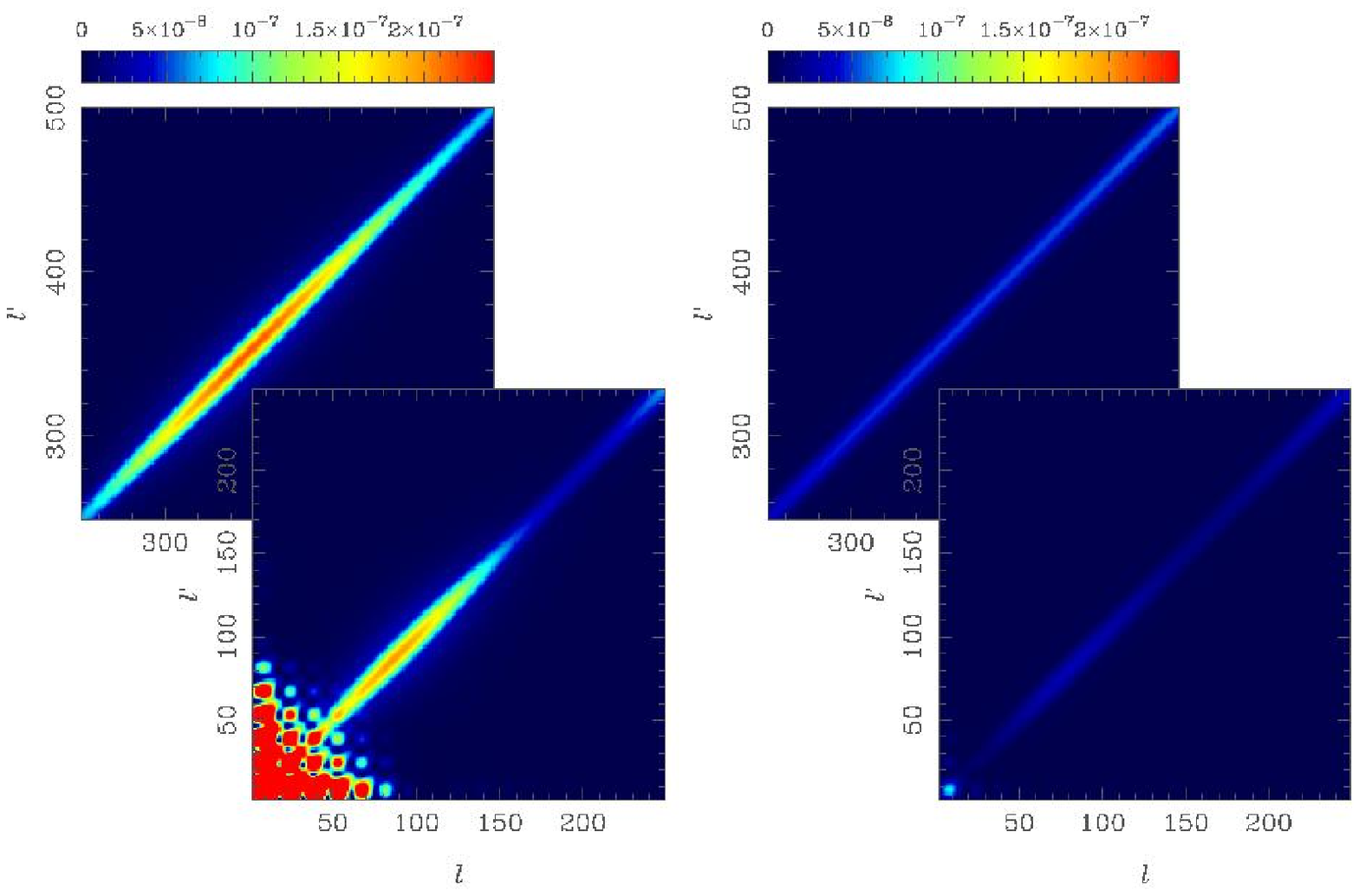,angle=0,width=16cm}
\caption{Blocks of the covariance matrix
$l(l+1)l'(l'+1)\rmn{cov}(\tilde{C}_l^B, \tilde{C}_{l'}^B)$ for
a $15\degr$-radius region with weighting as in Fig.~\ref{fig:power_and_window}.
The exact covariance matrix is shown on the left, and
the contribution from $B$ modes is shown on the right (i.e.\ with
$C_l^E$ set to zero).
\label{fig:exactBB_10_15_cosine_and_exactBB_10_15_cosine_Bonly}
}
\end{figure}

As our second example we consider the $15\degr$-radius survey with cosine
apodization over the last $5\degr$ that we used in
Fig.~\ref{fig:power_and_window}. We should expect the relative
importance of $E$-$B$ mixing to extend to higher $l$ for this smaller
survey, and this is indeed seen in
Fig.~\ref{fig:exactBB_10_15_cosine_and_exactBB_10_15_cosine_Bonly} where
we compare the exact $\rmn{cov}(\tilde{C}_l^B, \tilde{C}_{l'}^B)$ with
the covariance obtained by setting $C_l^E=0$. The $E$-mode power dominates
the covariance for all scales plotted (maximum $l=500$), and the
magnitude of the diagonal follows the acoustic peaks in $C_l^E$.
Surveys of this size are being considered by a number of upcoming experiments
targeting $B$-mode polarization from gravitational waves. It is clear that,
even for the relatively optimistic tensor-to-scalar ratio adopted here
($r=0.15$), the application of pseudo-$C_l$ techniques to such surveys
will demand careful modelling of the impact of $E$-$B$ mixing in the
covariance matrix for $\tilde{C}_l^B$ on any scales where sample variance
is expected to dominate.

\section{Approximate covariances for smooth weighting}
\label{sec:approx-cov}

Given that exact evaluation of the pseudo-$C_l$ covariance matrices
is, in general, prohibitively expensive, in this section we aim to derive
accurate analytic approximations that allow fast computation of the covariances
at high $l$. This problem has been well-studied for temperature
anisotropies~\citep{hinshaw03,chon04,efstathiou04}, but, as yet, there
has been no comparable treatment for polarization. As with the
temperature case, the main simplification arises from assuming that
we are working at sufficiently high $l$, and with a weight function
that is sufficiently band-limited, that we can approximate the
$C_l$s as being constant over the width of all coupling matrices.
This removes the $C_l$s from all convolutions.
For ${}_{+}I_{(lm)(lm)'}$, and its products, this
approximation should be good, as in the temperature case, for
smoothly-apodized observations covering a connected patch of linear extent more
than several degrees. The approximation is poorer for ${}_{-}
I_{(lm)(lm)'}$, and breaks down completely if there is no apodization,
i.e.~the weight function $w(\vnhat)$ is set to unity or
zero. (See~\citealt{lewis02} for examples of elements of ${}_{-}I_{(lm)(lm)'}$
in this case.) Here we restrict ourselves to situations where these assumptions
hold. We first consider a specific case of Gaussian weighting over small
patches of the sky, before considering general (but smooth) weighting
on the sphere in Section~\ref{subsec:general}.

\subsection{Gaussian weighting in the flat-sky limit}
\label{subsec:gaussian}

As a warm-up to the approximate calculation of the pseudo-$C_l$
covariances for general weighting functions $w(\vnhat)$, we first consider
Gaussian weighting over small patches of the sky. For such observations we
can approximate the spherical-harmonic analysis by Fourier analysis, and
we are able to make analytic progress more simply than for the general case.
The results of this subsection are also of considerable interest for
observations of CMB polarization with interferometers
[such as DASI~\citep{kovac02}, CBI~\citep{readhead04} and
the Array for Microwave Background Anisotropy (AMiBA)\footnote{%
http://amiba.asiaa/sinica.edu.tw/}],
since they directly sample the Fourier transform of
the sky multiplied by the primary
beam. The latter thus acts as a weight function $w(\vnhat)$ for an
interferometer. Furthermore, the primary beam can often be approximated as a
Gaussian.

In the flat-sky approximation, for lines of sight close to the $z$ axis,
we adopt a global polarization basis that is
aligned with the $x$ and $-y$ axes. The radiation propagation direction at
the centre of the field is along $-\hat{\bmath{z}}$.
Stokes parameters defined on this basis, $Q(\vx)$ and $U(\vx)$, at
position $\vx$ on the sky can be decomposed into Fourier modes
$Q(\vl)$ and $U(\vl)$:
\begin{equation}
Q(\vl) \equiv \int \frac{d^2 \vx}{2\pi}\, Q(\vx) e^{-i\vl\cdot \vx}, \quad
U(\vl) \equiv \int \frac{d^2 \vx}{2\pi}\, U(\vx) e^{-i\vl\cdot \vx}.
\label{eq:flat1}
\end{equation}
From these we can define the Fourier transforms of the electric and
magnetic parts of the polarization,
\begin{equation}
-(E  \mp i B)(\vl) = e^{\pm 2i \phi_{\vl}} (Q\pm i U)(\vl),
\label{eq:flat2}
\end{equation}
where $\phi_{\vl}$ is the angle between $\vl$ and the $x$ axis.
Note that this is a rotation of the Stokes parameters in Fourier space
onto a basis adapted to the wavevector $\vl$. The Fourier transforms
$E(\vl)$ and $B(\vl)$ can be shown to be related to the spherical multipoles
$E_{lm}$ and $B_{lm}$ for $l \gg 1$ by the same correspondence as for the
temperature anisotropies, i.e.\
\begin{equation}
E_{lm} \approx (-i)^m \sqrt{\frac{l}{2\pi}} \int \ud\phi_{\vl}\, E(\vl)
e^{-im\phi_{\vl}}.
\end{equation}
A similar relation holds for $B$. In terms of the power
spectra, we have e.g.\ $\langle E(\vl) E^*(\vl') \rangle = C_l^E \delta(\vl-
\vl')$.

If the polarization is weighted by $w(\vx)$, then the Fourier transform of
the Stokes parameters for the weighted field (or the visibilities in the
case of an interferometer) are
\begin{eqnarray}
\tilde{Q}(\vl) &=& \int \frac{d^2 \vx}{2\pi}\, w(\vx) Q(\vx)
e^{-i\vl\cdot \vx}, \label{eq:flat3} \\
	       &=& \int \frac{d^2 \vl'}{2\pi}\, w(\vl-\vl')Q(\vl'),
\label{eq:flat4}
\end{eqnarray}
with a similar expression for $\tilde{U}(\vl)$.
Here $w(\vl)$ is the Fourier transform of $w(\vx)$. For $w(\vx)$ real,
$\tilde{Q}^\ast(\vl)= \tilde{Q}(-\vl)$ and similarly for $\tilde{U}(\vl)$.
Decomposing $\tilde{Q}(\vl)$
and $\tilde{U}(\vl)$ as in equation~(\ref{eq:flat2}), we find
\begin{equation}
(-\tilde{E}\pm i \tilde{B})(\vl) = \int \frac{d^2 \vl'}{2\pi} \,
e^{\pm 2i (\phi_{\vl} - \phi_{\vl'})} w(\vl-\vl') (-E \pm i B)(\vl'),
\label{eq:flat5}
\end{equation}
or equivalently
\begin{eqnarray}
\tilde{E}(\vl) &=& \int \frac{d^2 \vl'}{2\pi} [{}_+I(\vl,\vl') E(\vl')
+ i {}_-I(\vl,\vl') B(\vl')] , \label{eq:flat6} \\
\tilde{B}(\vl) &=& \int \frac{d^2 \vl'}{2\pi} [{}_+I(\vl,\vl') B(\vl')
- i {}_-I(\vl,\vl') E(\vl')] , \label{eq:flat7}
\end{eqnarray}
where the Hermitian [for $w(\vx)$ real] kernels are
\begin{equation}
{}_+I(\vl,\vl') = w(\vl-\vl') \cos 2(\phi_{\vl}-\phi_{\vl'}),
\quad {}_-I(\vl,\vl') = -i w(\vl-\vl') \sin 2(\phi_{\vl}-\phi_{\vl'}).
\label{eq:flat8}
\end{equation}
These are the flat-sky limits of the ${}_\pm I_{(lm)(lm)'}$ in
equations~(\ref{eq:5-27}) and (\ref{eq:5-28}).
The geometric mixing of e.g.\ $E(\vl)$ into $\tilde{B}
(\vl)$ is suppressed by the trigonometric factor $\sin2(\phi_\vl -\phi_{\vl'})$.
If the extent of the support of $w(\vl)$ is $\sim l_{\rmn{max}}$ [i.e.\
the band-limit of $w(\vnhat)$] then the mixing of $E$ into the observable
$\tilde{B}$ will be $\sim (l_{\rmn{max}} / l) E(\vl)$ for
$l \gg l_{\rmn{max}}$. Clearly this is significant if
$|B(\vl)| \la (l_{\rmn{max}} / l) |E(\vl)|$.

\subsubsection{Pseudo-$C_l$ covariance}

In the flat-sky limit the pseudo-$C_l$s are defined by
\begin{equation}
\tilde{C}_l^E \equiv \frac{1}{2} \int d\phi_\vl \, |\tilde{E}(\vl)|^2, \quad
\tilde{C}_l^B \equiv \frac{1}{2} \int d\phi_\vl \, |\tilde{B}(\vl)|^2.
\label{eq:flat35}
\end{equation}
The signal correlations between the $\tilde{E}(\vl)$ and $\tilde{B}(\vl)$
are discussed in Appendix~\ref{app:flat}. For Gaussian weighting,
$w(\vx)=\exp(-\vx^2/2\sigma^2)$, and smooth spectra (over a range of multipoles
$\sim 1/\sigma$), we can use equations~(\ref{eq:flat18})--(\ref{eq:flat20})
to show that
\begin{equation}
\langle \tilde{C}_l^E \rangle \approx \alpha(l) C_l^E
+ \beta(l) C_l^B ,\quad
\langle \tilde{C}_l^B \rangle \approx \alpha(l) C_l^B
+ \beta(l) C_l^E, \label{eq:flat37}
\end{equation}
where
\begin{eqnarray}
\alpha(l) &=& \frac{\sigma^2}{2}\left[ \frac{1}{2} - \frac{1}{l^2\sigma^2}
\left(1-\frac{3}{2l^2\sigma^2}\right) - \frac{e^{-l^2\sigma^2}}{2l^2\sigma^2}
\left(1+\frac{3}{l^2\sigma^2}\right)\right], \label{eq:flat27} \\
\beta(l) &=& \frac{1}{2 l^2}\left[
1-\frac{3}{2l^2\sigma^2}+ \frac{1}{2}e^{-l^2\sigma^2}
\left(1+\frac{3}{l^2\sigma^2}\right)\right]. \label{eq:flat28}
\end{eqnarray}
These results are consistent with equations~(\ref{eq:5-35a}) and
(\ref{eq:5-35}), if we use $\sum_{l'} P_{ll'} \approx \alpha(l)$
and $\sum_{l'} M_{ll'} \approx \beta(l)$ for $l \gg 1$,
since the (spherical) power spectrum of the Gaussian weight function
$w_l \approx \pi \sigma^4 \exp(-l^2\sigma^2)$.

The sample covariances of the pseudo-$C_l$s are
given by the angular average of the squares of the correlators of
$\tilde{E}(\vl)$ and $\tilde{B}(\vl)$:
\begin{eqnarray}
\rmn{cov}(\tilde{C}_l^E,\tilde{C}_{l'}^E) &=& \frac{1}{2} \int d\phi_\vl
d \phi_{\vl'} \, |\langle \tilde{E}(\vl) \tilde{E}^\ast(\vl') \rangle |^2,
\label{eq:flat38} \\
\rmn{cov}(\tilde{C}_l^B,\tilde{C}_{l'}^B) &=& \frac{1}{2} \int d\phi_\vl
d \phi_{\vl'} \, |\langle \tilde{B}(\vl) \tilde{B}^\ast(\vl') \rangle |^2,
\label{eq:flat39} \\
\rmn{cov}(\tilde{C}_l^E,\tilde{C}_{l'}^B) &=& \frac{1}{2} \int d\phi_\vl
d \phi_{\vl'} \, |\langle \tilde{E}(\vl) \tilde{B}^\ast(\vl') \rangle |^2.
\label{eq:flat40}
\end{eqnarray}
For an azimuthally-symmetric weight function, the correlators in these
integrals are invariant under rigid rotations of $\vl$ and $\vl'$, so the
integration over one of $\phi_\vl$ and $\phi_{\vl'}$ is trivial.
To evaluate equations~(\ref{eq:flat38})--(\ref{eq:flat40}) for the
case of Gaussian weighting we make use of the results for the correlators given
in Appendix~\ref{app:flat}.
The analysis is rather involved so we simplify things
a little by considering only the limits $C_l^B=0$ and $C_l^E=0$. If we
consider the covariance of $\tilde{C}_l^E$ with itself, the case
$C_l^B=0$ describes the contribution to the covariance if there were no
$E$-$B$ mixing, while the case $C_l^E=0$ describes the contribution that
arises solely from mixing.

Consider $C_l^B=0$ first. Then, for azimuthally-symmetric $w(\vx)$, we have
e.g.\
\begin{equation}
\rmn{cov}(\tilde{C}_l^E,\tilde{C}_{l'}^E) \approx \frac{C_l^E C_{l'}^E}{2}
\int \frac{d \psi}{2\pi} \, \left|
\int \frac{d^2\vL}{2\pi}\, {}_+I(\vl,\vL) {}_+I^\ast(\vl',\vL) \right|^2,
\label{eq:flat41}  
\end{equation}
where $\psi \equiv \phi_{\vl'} - \phi_\vl$, and similar expressions hold
for the other covariances. We have not been able to
perform the integrals over $\psi$ exactly, but it is possible to make
progress in the limit of large $l$ and $l'$ ($\gg 1/\sigma$). In this limit,
we can ignore the terms with factor $\exp[-(l^2+{l'}^2)\sigma^2/2]$ in
equations~(\ref{eq:flat20}) and (\ref{eq:flat23}) in Appendix~\ref{app:flat}.
This leaves a factor of $\exp(-|\vl-\vl'|^2 \sigma^4/2)$ multiplying
slowly-varying functions of $\psi$. The exponential peaks sharply around
$\psi=0$ (reflecting the fact that $\tilde{E}(\vl)$ and $\tilde{B}(\vl')$
are only strongly correlated for $\vl$ within a radius $\sim 1/\sigma$ of
$\vl'$), so we can accurately handle the inverse powers of $|\vl+\vl'|$
and trigonometric terms in equations~(\ref{eq:flat20}) and (\ref{eq:flat23})
with an expansion in $\sin^2\psi$. To make this expansion it is useful to
note that
\begin{eqnarray}
\cos2(\phi_\vl+\phi_{\vl'}-2\phi_{\vl+\vl'}) &=& 1 - 2
\frac{(l^2-{l'}^2)^2}{|\vl + \vl'|^4} \sin^2\psi,
\label{eq:flat42} \\
\sin2(\phi_\vl+\phi_{\vl'}-2\phi_{\vl+\vl'}) &=& 2
\frac{(l^2-{l'}^2)}{|\vl + \vl'|^4} [(l^2+{l'}^2) \cos\psi + 2 ll'] \sin\psi.
\label{eq:flat43} 
\end{eqnarray}
Integrating the square of the expansions over $\psi$ then yields a series
of modified Bessel functions $I_{2n}(ll'\sigma^2)$. These series are
cumbersome, so we only give the leading-order results in the asymptotic
regime ($l$, $l' \gg 1/\sigma$), where we can replace the Bessel functions
with their asymptotic expansions:
\begin{eqnarray}
\rmn{cov}(\tilde{C}_l^E,\tilde{C}_{l'}^E) &\sim& \frac{\sigma^4}{8}
\frac{C_l^E C_{l'}^E}{\sqrt{2\pi ll'\sigma^2}} e^{-(l-l')^2\sigma^2/2},
\label{eq:flat44} \\
\rmn{cov}(\tilde{C}_l^B,\tilde{C}_{l'}^B) &\sim&
\frac{6}{(l+l')^4} \frac{C_l^E C_{l'}^E}{\sqrt{2\pi ll'\sigma^2}}
e^{-(l-l')^2\sigma^2/2},
\label{eq:flat45} \\
\rmn{cov}(\tilde{C}_l^E,\tilde{C}_{l'}^B) &\sim&
\frac{l\sigma^2}{2l' (l+l')^2}
\frac{C_l^E C_{l'}^E}{\sqrt{2\pi ll'\sigma^2}} e^{-(l-l')^2\sigma^2/2}
\quad (C_l^B = 0) .
\label{eq:flat46}
\end{eqnarray}
To obtain these results it is necessary to expand the
integrand in equation~(\ref{eq:flat41}) to $O(\sin^4\psi)$.
It is straightforward to verify that the covariances of the pseudo-$C_l$s
for the other limit, $C_l^E=0$, can be obtained by interchanging
$E$ and $B$ in these expressions.

To gain an indication of how the covariance of the pseudo-$C_l$s propagates
to estimates of the power spectra, we form simple estimators for
$C_l^E$ and $C_l^B$ that are local in $\tilde{C}_l^E$ and $\tilde{C}_l^B$
by solving equation~(\ref{eq:flat37}):
\begin{equation}
\hat{C}_l^E = \frac{1}{N(l)}[\alpha(l) \tilde{C}_l^E - \beta(l)
\tilde{C}_l^B], \quad \hat{C}_l^B = \frac{1}{N(l)}[\alpha(l) \tilde{C}_l^B
- \beta(l) \tilde{C}_l^E], \label{eq:flat30} 
\end{equation}
where the normalisation $N(l) \equiv \alpha^2(l)-\beta^2(l)$.
The covariance of these estimators then follows from the covariance
of the pseudo-$C_l$s which we derived above. To leading order, we find
\begin{equation}
\rmn{cov}(\hat{C}_l^B,\hat{C}_{l'}^B) = \frac{16 C_l^E C_{l'}^E}{\sqrt{2\pi
ll' \sigma^2}} e^{-(l-l')^2\sigma^2/2} \left[\frac{6}{(l+l')^4\sigma^4}
-\frac{1}{(l+l')^2\sigma^4}\left(\frac{l}{{l'}^3}+\frac{l'}{l^3}\right)
+ \frac{1}{2l^2 {l'}^2 \sigma^4} \right] \quad (C_l^B=0),
\label{eq:flat47}
\end{equation}
and
\begin{equation}
\rmn{cov}(\hat{C}_l^B,\hat{C}_{l'}^B) = \frac{2 C_l^B C_{l'}^B}{\sqrt{2\pi
ll' \sigma^2}} e^{-(l-l')^2\sigma^2/2} \quad (C_l^E=0).
\label{eq:flat48}
\end{equation}
Note that these expressions are symmetric in $l$ and $l'$ as
required. Equation~(\ref{eq:flat48}), which gives the
covariance of the estimated $B$-mode power if there were no $E$-$B$ mixing,
is the direct analogue of the asymptotic
expression for the temperature anisotropies, where the exact result
in the flat-sky limit (for smooth $C_l^T$) is
\begin{equation}
\rmn{cov}(\hat{C}_l^T,\hat{C}_{l'}^T) = 2 C_l^T C_{l'}^T
e^{-(l^2+{l'}^2)\sigma^2/2} I_0(ll'\sigma^2).
\label{eq:flat49}
\end{equation}
If we average the estimators into bands of width $\Delta l \gg 1/\sigma$ we
obtain quasi-uncorrelated estimates. The variances of these
band-powers follows by integrating equations~(\ref{eq:flat47})
and (\ref{eq:flat48}) over $l$ and $l'$ within the band. For wide bands, we
find the approximate results
\begin{eqnarray}
\rmn{var}(\hat{C}_l^B) &\approx& \frac{6 {C_l^E}^2}{(l\sigma)^5
\Delta l \sigma} \quad (C_l^B=0) , \label{eq:flat50} \\
\rmn{var}(\hat{C}_l^B) &\approx&
\frac{2{C_l^B}^2}{l\Delta l\sigma^2}
\quad (C_l^E = 0). \label{eq:flat51}
\end{eqnarray}
For the $C_l^E=0$ case, we recover the approximation given for temperature
by \citet{hivon02},
\begin{equation}
\rmn{var}(\hat{C}_l^T) \approx \frac{2 w^{(4)}{C_l^T}^2}{(2l+1)\Delta l
f_{\rmn{sky}} {w^{(2)}}^2},
\label{eq:hivon}
\end{equation}
noting that, for the case of Gaussian weighting considered
here, $w^{(2)} f_{\rmn{sky}} = \sigma^2/4$ and $w^{(4)}
f_{\rmn{sky}} = \sigma^2/8$. In Section~\ref{subsec:rot} we develop
more general rules of thumb for arbitrary (but smooth) weight functions;
these properly reduce to the expressions derived here for
Gaussian weighting in the flat-sky limit.

Our leading-order results show that, for Gaussian weighting, the ratio of the
cosmic variance contributions to $\hat{C}_l^B$ from $E$ and $B$ is
$\sim (C_l^E/C_l^B)^2 / (l\sigma)^4$ for $l \gg 1/\sigma$.
In Section~\ref{sec:implications} we consider the implications of this
residual variance from $E$-mode power in unbiased quadratic estimates of
$C_l^B$ for detecting gravitational waves via
$B$-mode polarization.

\subsection{Approximations for general weighting}
\label{subsec:general}

We now consider the case of general (but smooth) weight functions and
work on the spherical sky. The starting point is to approximate the
correlators of the pseudo-multipoles that appear squared in the pseudo-$C_l$
covariances, equations~(\ref{eq:5-37})--(\ref{eq:5-39}), by removing
the $C_l$s from the convolutions:
\begin{eqnarray}
\langle \tilde{E}_{lm} \tilde{E}_{(lm)'}^* \rangle
	&=& \sum_{LM} {}_{+}I_{(lm)(LM)}{}_{+}I_{(lm)' (LM)}^* C_L^E
	+ {}_{-}I_{(lm)(LM)}{}_{-}
	I_{(lm)' (LM)}^* C_L^B
	\nonumber \\
	&\approx& \sqrt{C_l^E C_{l'}^E} [{}_{+}I(w^2)]_{(lm)(lm)'}
	- \left( \sqrt{C_l^E C_{l'}^E} - \sqrt{C_l^B C_{l'}^B} \right)
	[{}_{-}I^2(w)]_{(lm)(lm)'}, \label{eq:5-40}\\
\langle \tilde{B}_{lm} \tilde{B}_{(lm)'}^* \rangle
	&=& \sum_{LM} {}_{+}I_{(lm)(LM)}{}_{+}I_{(lm)' (LM)}^* C_L^B
	+ {}_{-}I_{(lm)(LM)}{}_{-} I_{(lm)' (LM)}^* C_L^E 
	\nonumber \\
	&\approx& \sqrt{C_l^B C_{l'}^B} [{}_{+}I(w^2)]_{(lm)(lm)'}
	+ \left( \sqrt{C_l^E C_{l'}^E} - \sqrt{C_l^B C_{l'}^B} \right)
	[{}_{-}I^2(w)]_{(lm)(lm)'}, \label{eq:5-41}\\
\langle \tilde{E}_{lm} \tilde{B}_{(lm)'}^* \rangle
	&=& i \sum_{LM} {}_{+}I_{(lm)(LM)}{}_{-}I_{(lm)' (LM)}^* C_L^E
	+ {}_{-}I_{(lm)(LM)}{}_{+} I_{(lm)' (LM)}^* C_L^B \nonumber \\
&\approx&  \frac{i}{2} \left(\sqrt{C_l^E C_{l'}^E}+
	\sqrt{C_l^B	C_{l'}^B}\right)
	[{}_{-}I(w^2)]_{(lm)(lm)'} \nonumber \\
&& \mbox{} + \frac{i}{2}
	\left(\sqrt{C_l^E C_{l'}^E}-\sqrt{C_l^B C_{l'}^B}\right)
	\left[{}_{+}I(w){}_{-}I(w)- {}_{-}I(w){}_{+}I(w)\right]_{(lm)(lm)'},
\label{eq:5-42}
\end{eqnarray}
where $[{}_{\pm}I(w^2)]_{(lm)(lm)'}$ are the matrices ${}_{\pm}
I_{(lm)(lm)'}$ but evaluated with weight function $w^2(\vnhat)$ rather than
$w(\vnhat)$, and $[{}_- I^2(w)]_{(lm)(lm)'}$ is the matrix product
$\sum_{LM} {}_{-}I_{(lm)(LM)} {}_{-}I_{(LM)(lm)'}$.\footnote{Due to
the Hermiticity of ${}_\pm I_{(lm)(lm)'}$, $[{}_- I^2(w)]_{(lm)(lm)'}$ is
also equal to $\sum_{LM}{}_{-}I_{(lm)(LM)} {}_{-}I^*_{(lm)'(LM)}$.}
We now consider the derivation of equations~(\ref{eq:5-40})--(\ref{eq:5-42})
in detail. We start from the completeness relation $\sum_{lm}
{}_{s} Y_{lm} (\vnhat) {}_{s} Y^*_{lm} (\vnhat')=\delta(\vnhat-\vnhat')$ which
ensures that 
\begin{eqnarray}
\left[{}_{\pm 2} I^2(w)\right]_{(lm)(lm)'} &=& \sum_{LM} {}_{\pm 2}I_{(lm)(LM)}
{}_{\pm 2} I^*_{(lm)' (LM)} \nonumber\\
&=& \int \ud \vnhat \; w^2 (\vnhat) {}_{\pm 2} Y_{lm}^* (\vnhat) 
{}_{\pm 2} Y_{(lm)'}(\vnhat)
\nonumber \\
&\equiv& [{}_{\pm 2} I(w^2)]_{(lm)(lm)'}.
\label{eq:5-43}
\end{eqnarray}
For $w(\vnhat)=1$ or 0, so that $w^2
(\vnhat)=w(\vnhat)$, these relations imply that ${}_{\pm 2} I_{(lm)(lm)'}$
are projection operators~\citep{lewis02}. If we express ${}_{\pm 2} I$ in terms
of ${}_{\pm} I$, using the obvious matrix notation, we find 
\begin{equation}
{}_{\pm 2} I(w^2) = {}_{+}I^2 (w) + {}_{-}I^2 (w) \pm {}_{+}I(w) {}_{-}I(w) \pm {}_{-}
I(w) {}_{+}I(w).
\label{eq:5-44}
\end{equation}
Adding and subtracting gives 
\begin{equation}
{}_{+}I^2 (w) + {}_{-}I^2 (w) = {}_{+}I(w^2), \quad
{}_{+}I (w) {}_{-}I(w) + {}_{-}I(w) {}_{+}I(w) = {}_{-}I(w^2),
\label{eq:5-45}
\end{equation}
which can be used to verify the second equalities in
equations~(\ref{eq:5-40})--(\ref{eq:5-42}). Note that if $C_l^E$ were
equal to $C_l^B$, the correlators would only involve ${}_{\pm} I(w^2)$ and
the construction of the covariances would then be no more difficult than
for the temperature anisotropies. The major simplification that arises in
this case is that the correlators of the pseudo-multipoles reduce to
linear functionals of $w^2(\vnhat)$. For statistically-isotropic CMB signals,
the pseudo-$C_l$ covariance must be invariant under rotations of
$w(\vnhat)$, so that if the correlators are linear in $w^2(\vnhat)$
then the covariance matrix can only depend on rotationally-invariant
quadratic combinations of the multipoles of $w^2$, i.e.\ the power spectrum
of $w^2$. In the general case, where the pseudo-multipole correlators
are more-general quadratic functionals of $w(\vnhat)$, the covariance matrix
can depend on less-specific configurations of the trispectrum of $w(\vnhat)$.
(The power spectrum of $w^2(\vnhat)$ picks up only rather specific -- and
easy to calculate -- configurations of the full trispectrum.)

Forming the covariances from
equations~(\ref{eq:5-37})--(\ref{eq:5-39}), we now find
\begin{eqnarray}
{\rmn{cov}} \left(\tilde{C}_l^E,\tilde{C}_{l'}^E\right) 
&\approx& \frac{2}{(2l+1)(2l'+1)} \sum_{mm'} \left|\sqrt{C_l^E C_{l'}^E}
[{}_{+}I(w^2)]_{(lm)(lm)'}
- \left( \sqrt{C_l^E C_{l'}^E} - \sqrt{C_l^B C_{l'}^B} \right)
[{}_{-}I^2(w)]_{(lm)(lm)'}\right|^2, 
\label{eq:5-46}\\
{\rmn{cov}} \left(\tilde{C}_l^B,\tilde{C}_{l'}^B\right)
&\approx& \frac{2}{(2l+1)(2l'+1)} \sum_{mm'} \left|\sqrt{C_l^B C_{l'}^B}
[{}_{+}I(w^2)]_{(lm)(lm)'}
- \left( \sqrt{C_l^B C_{l'}^B} - \sqrt{C_l^E C_{l'}^E} \right)
[{}_{-}I^2(w)]_{(lm)(lm)'}\right|^2, 
\label{eq:5-47}\\
{\rmn{cov}} \left(\tilde{C}_l^E,\tilde{C}_{l'}^B\right) 
&\approx& \frac{2}{(2l+1)(2l'+1)} \sum_{mm'} \left|
\frac{i}{2} \left(\sqrt{C_l^E C_{l'}^E}+\sqrt{C_l^B C_{l'}^B}\right)
[{}_{-}I(w^2)]_{(lm)(lm)'} \right. \nonumber \\
&& \hspace{0.2\textwidth} \mbox{} \left.
+ \frac{i}{2} \left(\sqrt{C_l^E C_{l'}^E}-\sqrt{C_l^B
C_{l'}^B}\right) 
\left[{}_{+}I(w){}_{-}I(w)- {}_{-}I(w){}_{+}I(w)\right]_{(lm)(lm)'} \right|^2.
\label{eq:5-48}
\end{eqnarray}
It is straightforward to compute ${}_{\pm} I(w^2)$ in terms of Wigner-3$j$
symbols using equation~(\ref{eq:5-26}):
\begin{equation}
[{}_{\pm} I(w^2)]_{(lm)(lm)'} = \frac{1}{2} \sum_{LM} (-1)^m (w^2)_{LM}
\sqrt{\frac{(2l+1)(2l'+1)(2L+1)}{4\pi}} [1\pm (-1)^{K}]
\left( \begin{array}{ccc} l & l' & L \\ -2 & 2 & 0
\end{array} \right) \left( \begin{array}{ccc} l & l' & L \\ m & -m' & -M
\end{array} \right),
\label{eq:5-49}
\end{equation}
where $(w^2)_{LM}$ are the (scalar) multipoles of $w^2(\vnhat)$,
and, as before, $K\equiv l+l'+L$. For ${}_{-}I^2(w)$,
we go back one step, taking as our starting point the integral expression for
${}_{-}I_{(lm)(lm)'}$:
\begin{eqnarray}
{}_{-}I_{(lm)(lm)'} &=& \int \ud\vnhat \; w(\vnhat) \frac{1}{2}
\left[ {}_{2}Y_{lm}^*(\vnhat) {}_{2} Y_{(lm)'}(\vnhat)
- {}_{-2}Y_{lm}^*(\vnhat) {}_{-2}Y_{(lm)'}(\vnhat) \right].
\label{eq:5-50}
\end{eqnarray}
Our aim is to find an approximation to ${}_{-}I^2 (w)$ that is a
functional of the squares of derivatives of $w$, so that its contribution
to the covariance matrices will only depend on easily-calculable
configurations of the trispectrum of $w$. Writing the spin-2 harmonics in
equation~(\ref{eq:5-50}) in terms of spin-weight derivatives of the
scalar harmonics (see e.g.\ Appendix B of~\citealt{lewis02}),
\begin{eqnarray}
{}_{2}Y_{lm} &=& \sqrt{\frac{(l-2)!}{(l+2)!}} \eth^2 Y_{lm}, \qquad
{}_{-2}Y_{lm} = \sqrt{\frac{(l-2)!}{(l+2)!}} \mybeth^2 Y_{lm},
\label{eq:5-51}
\end{eqnarray}
which follow from repeated use of
\begin{equation}
\eth {}_s Y_{lm} = \sqrt{(l-s)(l+s+1)} {}_{s+1} Y_{lm}, \quad 
\mybeth {}_s Y_{lm} = - \sqrt{(l+s)(l-s+1)} {}_{s-1} Y_{lm},
\label{eq:5-52}
\end{equation}
integrating by parts and using 
$(\eth^2\mybeth^2-\mybeth^2\eth^2)Y_{lm}=0$, we find that
\begin{equation}
{}_{-}I_{(lm)(lm)'} = \frac{1}{2} \sqrt{\frac{(l'-2)!}{(l'+2)!}} \int
\ud\vnhat \, \left[ \eth^2 w(\vnhat) {}_{2}Y_{lm}^*(\vnhat) - \mybeth^2
w(\vnhat) {}_{-2} Y_{lm}^*(\vnhat)
+ 2 \eth w(\vnhat) \eth {}_{2}Y_{lm}^*(\vnhat) - 2 \mybeth
w(\vnhat)\mybeth {}_{-2}Y_{lm}^*(\vnhat)\right] Y_{(lm)'}(\vnhat). 
\label{eq:5-53}
\end{equation}
This result shows explicitly that ${}_{-}I_{(lm)(lm)'}$ vanishes if
$w(\vnhat)$ is a constant over the full sky. More generally, for
$w(\vnhat)$ band-limited to $l_{\rmn{max}}$, it suggests that ${}_{-}
I_{(lm)(lm)'} \sim O(l_{\rmn{max}}/l) {}_+ I_{(lm)(lm)'}$
for $l\gg l_{\rmn{max}}$ with this
leading-order contribution coming from the last two terms in
equation~(\ref{eq:5-53}). Again, this is consistent with our expectation
that $E$-$B$ mixing is suppressed for $l \gg l_{\rmn{max}}$.
Forming the square of ${}_{-}I_{(lm)(lm)'}$, we find
\begin{eqnarray}
[{}_{-}I^2(w)]_{(lm)(lm)'}
&=&\frac{1}{4} \int\ud\vnhat_1\ud\vnhat_2\, 
\Biggl\{ \left[ \eth^2 w(\vnhat_1) {}_{2}Y_{lm}^*(\vnhat_1) - \mybeth^2
w(\vnhat_1) {}_{-2}Y_{lm}^*(\vnhat_1) + 2 \eth w(\vnhat_1) \eth
{}_{2}Y_{lm}^*(\vnhat_1) - 2 \mybeth w(\vnhat_1) \mybeth
{}_{-2}Y_{lm}^*(\vnhat_1)
\right]
\nonumber \\ && \times
\left.\left[ \eth^2 w(\vnhat_2) {}_{2}Y_{(lm)'}^*(\vnhat_2) -\mybeth^2
w(\vnhat_2) {}_{-2}Y_{(lm)'}^*(\vnhat_2) + 2 \eth 
w(\vnhat_2) \eth {}_{2}Y_{(lm)'}^*(\vnhat_2) - 2 \mybeth w(\vnhat_2)
\mybeth {}_{-2} Y_{(lm)'}^*(\vnhat_2) \right]^* \right.
\nonumber \\ && \times \left.
\sum_{L\ge 2,M} Y_{LM}(\vnhat_1) Y_{LM}^*(\vnhat_2) \frac{(L-2)!}{(L+2)!} 
\right\}.
\label{eq:5-54}
\end{eqnarray}
Although we shall not make use of it here, we note for completeness that
the sum over $L$ in this equation can be performed exactly as follows:
(i) apply the operator $\nabla^2 (\nabla^2+2)$ to remove the factor
$(L-2)!/(L+2)!$; 
(ii) perform the resulting summation with the addition theorem; (iii)
solve the differential equation away from $\vnhat_1 \cdot \vnhat_2 = 1$;
and (iv) enforce the correct behaviour at $\vnhat_1 \cdot \vnhat_2 = 1$.
The final result is
\begin{equation}
\sum_{L\ge 2,M} Y_{LM}(\vnhat_1) Y_{LM}^*(\vnhat_2) \frac{(L-2)!}{(L+2)!}
= \frac{1-2\ln 2}{16\pi} + \frac{1+6\ln 2}{48\pi}\cos\beta
+ \frac{1}{8\pi} (1-\cos\beta)\ln(1-\cos\beta),
\label{eq:gen1}
\end{equation}
where $\cos\beta \equiv \vnhat_1 \cdot \vnhat_2$. The first two terms
can be shown not to contribute to ${}_-I^2(w)$. For our purposes, it is
more useful to approximate the sum in equation~(\ref{eq:5-54}), for $l,l' \gg
l_{\rmn{max}}$, by 
\begin{eqnarray}
\sum_{L\ge 2, M} \frac{(L-2)!}{(L+2)!} Y_{LM}(\vnhat_1)
Y_{LM}^*(\vnhat_2) &\approx& \sqrt{\frac{(l-2)!}{(l+2)!}}
\sqrt{\frac{(l'-2)!}{(l'+2)!}}
\sum_{L\ge 0, M} Y_{LM}(\vnhat_1) Y_{LM}^*(\vnhat_2)
\nonumber \\
&=& \sqrt{\frac{(l-2)!}{(l+2)!}} \sqrt{\frac{(l'-2)!}{(l'+2)!}}
\delta(\vnhat_1 -\vnhat_2).
\label{eq:5-55}
\end{eqnarray}
Note that this expression only makes sense when considered \emph{inside the
integrals} in equation~(\ref{eq:5-54}). The justification for the
approximation is as follows.
Equation~(\ref{eq:5-54}) may be regarded as a convolution of the
slowly-varying $(L-2)!/(L+2)!$  with the product of two functions of
$L$: one resulting from the integral over $\vnhat_1$, which is
localised to within $l_{\rmn{max}}$ of $l$, and the other from
the integral over $\vnhat_2$, which is within $l_{\rmn{max}}$ of $l'$.
We can thus replace $(L-2)!/(L+2)!$ with $\sqrt{(l-2)!/(l+2)!}\sqrt{(l'-2)!/
(l'+2)!}$ in the convolution for $l$ and $l'\gg l_{\rmn{max}}$.
Furthermore, we can then extend the summation to include
the $L=0$ and $L=1$ terms since they sum to a constant plus $\cos\beta$ which,
like the first two terms on the right of equation~(\ref{eq:gen1}),
do not contribute to the
integral. (These terms would be negligible anyway, for large $l$ and $l'$.)
The presence of the delta function in equation~(\ref{eq:5-55}) ensures that we
achieve our aim of reducing ${}_- I^2(w)$ to a linear functional of
squares of derivatives of $w(\vnhat)$. However, before proceeding we make
one further approximation which can easily be relaxed if more accuracy is
required. Since $w(\vnhat)$ is assumed to be slowly
varying compared to the spherical harmonics, we ignore terms like $\eth^2 w
{}_{2}Y_{lm}^*$ compared to e.g.~$\eth w \eth {}_{2} Y^*_{lm}$, so that we
retain only the leading-order contribution to ${}_- I^2(w)$. Finally, with
these approximations we have
\begin{eqnarray}
[{}_{-}I^2(w)]_{(lm)(lm)'} 
&\approx&
\sqrt{\frac{(l-2)!}{(l+2)!}}\sqrt{\frac{(l'-2)!}{(l'+2)!}}
\int \ud \vnhat \left\{ \left[ \eth w(\vnhat) \eth {}_{2}Y_{lm}^*(\vnhat) -
\mybeth  w(\vnhat) \mybeth\! {}_{-2}Y_{lm}^*(\vnhat) \right] \right.
\nonumber \\
&& \mbox{} \left. \times \left[ \eth  w(\vnhat) \eth {}_{2}Y_{(lm)'}^*(\vnhat)
- \mybeth w(\vnhat) \mybeth {}_{-2}Y_{(lm)'}^*(\vnhat) \right]^* \right\}
 \nonumber \\
&=& \frac{(-1)^{m+1}}{\sqrt{l(l+1)l'(l'+1)}}\sum_{LM}
\sqrt{\frac{(2l+1)(2l'+1)(2L+1)}{4\pi}}
\left( \begin{array}{ccc} l & l' & L \\ m & -m' & -M \end{array} \right) 
\nonumber \\ && \times
\left\{ \left| \eth w \right|^2_{LM} [1+(-1)^{K}]
\left( \begin{array}{ccc} l & l' & L \\ -1 & 1 & 0 \end{array} \right) 
+ \left[ {}_{2}(\eth  w)^2_{LM} +(-1)^{K} {}_{-2}(\mybeth w)^2_{LM}
\right] 
\left( \begin{array}{ccc} l & l' & L \\ -1 & -1 & 2 \end{array} \right) 
\right\},
\label{eq:5-56}
\end{eqnarray}
where we have used equation~(\ref{eq:5-52}). Note that this approximation
preserves Hermiticity of ${}_- I^2(w)$. In deriving equation~(\ref{eq:5-56})
we have expanded
products of the derivatives of $w(\vnhat)$ in the appropriate
spin-weight harmonics:
\begin{eqnarray}
\eth w(\vnhat) [\eth w(\vnhat)]^* 
&=& \mybeth w(\vnhat) [\mybeth  w(\vnhat)]^*
= |\eth w(\vnhat)|^2 
= \sum_{l\ge 0,\,m} |\eth w|^2_{lm} Y_{lm}(\vnhat),
\\
\eth w(\vnhat) [\mybeth w(\vnhat)]^* 
&=& [\eth w(\vnhat)]^2
= \sum_{l\ge 2,\,m} {}_{2}(\eth  w)^2_{lm} {}_{2}Y_{lm}(\vnhat),
\\
\mybeth  w(\vnhat) [\eth  w(\vnhat)]^* 
&=& [\mybeth  w(\vnhat)]^2
= \sum_{l\ge 2,\,m} {}_{-2}(\mybeth  w)^2_{lm} {}_{-2}Y_{lm}(\vnhat).
\label{eq:5-57}
\end{eqnarray}
Note that $\mybeth w(\vnhat)$ is the complex conjugate of
$\eth w(\vnhat)$ since $w$ is real, and that the spin-2 multipoles satisfy
${}_{-2}(\mybeth w)^2_{lm}= (-1)^m {}_{2}(\eth w)^{2*}_{l-m}$. Substituting
equations~(\ref{eq:5-49}) and~(\ref{eq:5-56}) into equations~(\ref{eq:5-46})
and~(\ref{eq:5-47}), we have 
\begin{eqnarray}
{\rmn{cov}} \left( \tilde{C}_l^E, \tilde{C}_{l'}^E \right)
&=&  \frac{1}{4\pi} \sum_{LM} \Biggr\{ [1+(-1)^{K}]
\left| \sqrt{C_l^E C_{l'}^E} (w^2)_{LM}
\left( \begin{array}{ccc} l & l' & L \\ -2 & 2 & 0 \end{array} \right)
\nonumber \right. \\
&&\mbox{} \left.
+ \frac{2\left(\sqrt{C_l^E C_{l'}^E}-\sqrt{C_l^B C_{l'}^B}\right)}
{\sqrt{l(l+1)l'(l'+1)}} \left[ |\eth w|^2_{LM}
\left( \begin{array}{ccc} l & l' & L \\ -1 & 1 & 0 \end{array} \right) 
+ \cle_{LM}
\left( \begin{array}{ccc} l & l' & L \\ -1 & -1 & 2 \end{array} \right) 
\right] \right|^2 \Biggr\} \nonumber \\
&&\mbox{} +
\frac{1}{\pi} \left[ \frac{\sqrt{C_l^E C_{l'}^E}-\sqrt{C_l^B
C_{l'}^B}}{\sqrt{l(l+1)l'(l'+1)}}\right]^2
\sum_{LM} [1-(-1)^{K}] |\clb_{LM}|^2
\left( \begin{array}{ccc} l & l' & L \\ -1 & -1 & 2 \end{array} \right)^2, 
\label{eq:5-58}
\end{eqnarray}
and
\begin{eqnarray}
{\rmn{cov}} \left( C_l^B, \tilde{C}_{l'}^B \right)
&=& \frac{1}{4\pi} \sum_{LM} \Biggr\{ [1+(-1)^{K}]
\left|  \sqrt{C_l^B C_{l'}^B} (w^2)_{LM}
\left( \begin{array}{ccc} l & l' & L \\ -2 & 2 & 0 \end{array} \right)
\right. \nonumber \\
&&\mbox{}\left.
+ \frac{2\left(\sqrt{C_l^B C_{l'}^B}-\sqrt{C_l^E C_{l'}^E}\right)}
{\sqrt{l(l+1)l'(l'+1)}} \left[ |\eth w|^2_{LM}
\left( \begin{array}{ccc} l & l' & L \\ -1 & 1 & 0 \end{array} \right) 
+ \cle_{LM}
\left( \begin{array}{ccc} l & l' & L \\ -1 & -1 & 2 \end{array} \right) 
\right] \right|^2 \Biggr\} \nonumber \\
&&\mbox{}
+ \frac{1}{\pi} \left[ \frac{\sqrt{C_l^B C_{l'}^B}-\sqrt{C_l^E
C_{l'}^E}}{\sqrt{l(l+1)l'(l'+1)}} \right]^2
\sum_{LM} [1-(-1)^{K}] |\clb_{LM}|^2
\left( \begin{array}{ccc} l & l' & L \\ -1 & -1 & 2 \end{array} \right)^2,
\label{eq:5-59}
\end{eqnarray}
where we have introduced parity eigenstates $\cle_{lm}$ and $\clb_{lm}$ for
the multipoles of the spin-$\pm2$ objects $[\eth w(\vnhat)]^2$ and
$[\mybeth w(\vnhat)]^2$, which are defined by\footnote{Under a parity
  transformation $w(\vnhat)\rightarrow w(-\vnhat)$, ${}_{2}(\eth 
  w)^2_{lm} \rightarrow (-1)^{l} {}_{- 2}(\mybeth w)^2_{lm}$ and ${}_{-2}
  (\mybeth  w)^2_{lm} \rightarrow (-1)^l {}_{2}(\eth w)^2_{lm}$.}
\begin{equation}
{}_{2}(\eth w)^2_{lm} = \cle_{lm}+i\clb_{lm},
\qquad
{}_{-2}(\mybeth w)^2_{lm} = \cle_{lm}-i\clb_{lm}.
\label{eq:5-60}
\end{equation}
Note that the $\clb_{lm}$ vanish for an azimuthally-symmetric weight
function since then ${}_{2}(\eth w)^2_{l0} = {}_{-2}(\mybeth w)^2_{l0}$.
Equations~(\ref{eq:5-58}) and~(\ref{eq:5-59}) achieve our
goal of finding approximations to the pseudo-$C_l$ covariance matrices
that are efficient to compute, but do take account of $E$-$B$ mixing
effects at leading order for $l$, $l'\gg l_{\rmn{max}}$.
Note that, whereas the pseudo-$C_l$ covariance for temperature anisotropies
involves only the power spectrum of $w^2$ (e.g.~\citealt{efstathiou04}),
the polarization covariances also
include the auto and cross-power (with $w^2$) of the square of the
gradient of $w$. 

We can make a useful check on our approximation for
$[{}_- I^2(w)]_{(lm)(lm)'}$, equation~(\ref{eq:5-56}), by setting
$l=l'$ and $m=m'$ and summing on $m$. Inspection of
equation~(\ref{eq:5-35}) shows that
\begin{equation}
\frac{1}{2l+1} \sum_m [{}_- I^2(w)]_{(lm)(lm)} = \sum_{l'} M_{ll'},
\end{equation}
whereas evaluating the left-hand side by summing the approximation
for ${}_- I^2(w)$ directly gives
\begin{equation}
\frac{1}{2l+1} \sum_m [{}_- I^2(w)]_{(lm)(lm)} \approx \frac{1}{\sqrt{\pi}
l(l+1)}(|\eth w|^2)_{00} = \frac{1}{2\pi} \int \frac{|\nabla w|^2}{l(l+1)}
\, \ud \vnhat.
\end{equation}
Reassuringly, this agrees with the leading-order result for
$\sum_{l'} M_{ll'}$ given in equation~(\ref{eq:15}).

Finally we consider the covariance of $\tilde{C}_l^E$ and $\tilde{C}_{l'}^B$,
equation~(\ref{eq:5-48}). This involves evaluating the matrix commutator
$[{}_{+}I, {}_{-}I]$. Following the procedure that led to
equation~(\ref{eq:5-53}) for ${}_{-}I_{(lm)(lm)'}$, we can express
${}_{+}I_{(lm)(lm)'}$ as
\begin{eqnarray}
{}_{+}I_{(lm)(lm)'} &=& \sqrt{\frac{(l'-2)!}{(l'+2)!}} \int \ud \vnhat 
\left\{ \frac{1}{2} \left[\eth^2 w(\vnhat) {}_{2}Y_{lm}^*(\vnhat) + \mybeth^2
w(\vnhat) {}_{-2}Y_{lm}^*(\vnhat)\right] +
\left[\eth w(\vnhat) \eth {}_{2}Y_{lm}^*(\vnhat) + \mybeth w(\vnhat)
\mybeth {}_{-2}Y_{lm}^*(\vnhat)\right] \right. \nonumber \\
&& \mbox{}\left.
+ \sqrt{\frac{(l+2)!}{(l-2)!}}w(\vnhat)Y_{lm}^*(\vnhat) \right\}
Y_{(lm)'}(\vnhat). 
\label{eq:5-61}
\end{eqnarray}
This is similar to equation~(\ref{eq:5-53}) except for the presence of
the last term in the curly brackets (and the sign differences).
The last term will dominate the
integral for $l$, $l'\gg l_{\rmn{max}}$, and in this limit
${}_{+}I_{(lm)(lm)'}$
will be close to the equivalent coupling matrix for the temperature
anisotropies (see e.g.~\citealt{hivon02}).
Forming the matrix product ${}_+ I(w) {}_- I(w)$ from
equations~(\ref{eq:5-53}) and (\ref{eq:5-61}), and using the
approximation in equation~(\ref{eq:5-55}), we find at leading order
\begin{eqnarray}
{}_{+}I(w) {}_{-}I(w) &\approx&
(-1)^m \sum_{LM} \sqrt{\frac{(2l+1)(2l'+1)(2L+1)}{4\pi}}
\left( \begin{array}{ccc} l & l' & L \\ m & -m' & -M \end{array} \right)
\nonumber \\
&& \mbox{}
\times \left[ \frac{1}{2} \sqrt{\frac{L(L+1)}{l'(l'+1)}} (w^2)_{LM} [1-(-1)^K]
\left( \begin{array}{ccc} l & l' & L \\ 0 & 1 & -1 \end{array} \right)
\right],
\label{eq:5-62}
\end{eqnarray}
which is $O(l_{\rmn{max}}/l)$ as expected. In deriving equation~(\ref{eq:5-62})
we have used $(w \eth w)=\eth w^2 / 2 = \sum_{LM}
\sqrt{L(L+1)} (w^2)_{LM}{}_{1}Y_{LM}/2$. Noting that ${}_{-}I(w) {}_{+}I(w)$ is
the Hermitian conjugate of ${}_{+}I(w) {}_{-}I(w)$, we have
\begin{eqnarray}
{}_{-}I(w) {}_{+}I(w) &\approx&
(-1)^{m+1} \sum_{LM} \sqrt{\frac{(2l+1)(2l'+1)(2L+1)}{4\pi}}
\left( \begin{array}{ccc} l & l' & L \\ m & -m' & -M \end{array} \right)
\nonumber \\
&& \mbox{} \times
\left[ \frac{1}{2} \sqrt{\frac{L(L+1)}{l(l+1)}} (w^2)_{LM} [1-(-1)^K]
\left( \begin{array}{ccc} l & l' & L \\ 1 & 0 & -1 \end{array} \right)
\right],
\label{eq:5-63}
\end{eqnarray}
where we have used the symmetries of the $3j$ symbols and the reality
of $w^2$. Differencing the previous two equations gives
\begin{eqnarray}
{}_{+}I(w) {}_{-}I(w) - {}_{-}I(w) {}_{+}I(w) 
&=& \sum_{LM} (-1)^m \sqrt{\frac{(2l+1)(2l'+1)(2L+1)}{4\pi}}
\left( \begin{array}{ccc} l & l' & L \\ m & -m' & -M \end{array} \right)
\frac{1}{2} \sqrt{L(L+1)} (w^2)_{LM} \nonumber \\
&& \mbox{}
\times  [1-(-1)^K] \left[  \frac{1}{\sqrt{l'(l'+1)}}
\left( \begin{array}{ccc} l & l' & L \\ 1 & 0 & -1 \end{array} \right)
+ \frac{1}{\sqrt{l(l+1)}} \left(
\begin{array}{ccc} l & l' & L \\ 0 & 1 & -1 \end{array} \right) \right] .
\label{eq:5-64}
\end{eqnarray}
This can be simplified further by noting that
\begin{equation}
[1-(-1)^K]\left[  \frac{1}{\sqrt{l'(l'+1)}}
\left( \begin{array}{ccc} l & l' & L \\ 1 & 0 & -1 \end{array} \right)
+ \frac{1}{\sqrt{l(l+1)}} \left(
\begin{array}{ccc} l & l' & L \\ 0 & 1 & -1 \end{array} \right) \right] =
\frac{[1-(-1)^K]}{\sqrt{l'(l'+1)}}
\left( \begin{array}{ccc} l & l' & L \\ 0 & 1 & -1 \end{array} \right)
\left( 1-\frac{l'(l'+1)}{l(l+1)} \right),
\end{equation}
which follows from a recursion relation for the $3j$ symbols [equation (4)
of Section 8.6.2 of~\citet{varshalovich}]. Since
$|l-l'| \le l_{\rmn{max}}$, it follows that the commutator
$[{}_{+}I(w),{}_{-}I(w)]_{(lm)(lm)'}$ is $O(l_{\rmn{max}}/l)^2$, and
so we can safely neglect its contribution compared 
to that from ${}_{-}I(w^2)$ in equation~(\ref{eq:5-48}) at high $l$.
(The power spectrum prefactors are approximately equal for both terms.)
In this limit,
$\rmn{cov}(\tilde{C}_l^E,\tilde{C}_{l'}^B)$ is symmetric.
Therefore our final approximation
for the covariance of $E$ and $B$ is, after summing over $m$ and $m'$,
\begin{equation}
{\rmn{cov}} \left( \tilde{C}_l^E ,\tilde{C}_{l'}^B \right)
\approx \frac{1}{16\pi} \sum_{LM} [1-(-1)^K] \left( \sqrt{C_l^E C_{l'}^E} +
\sqrt{C_l^B C_{l'}^B}\right)^2 |(w^2)_{LM}|^2 \left(
\begin{array}{ccc} l & l' & L \\ -2 & 2 & 0 \end{array} \right)^2.
\label{eq:5-65}
\end{equation}
This, and our approximations for $\rmn{cov}(\tilde{C}_l^E,
\tilde{C}_{l'}^E)$ and $\rmn{cov}(\tilde{C}_l^B,\tilde{C}_{l'}^B)$,
can be computed efficiently. For a general weight function $w(\vnhat)$,
we can compute the multipoles $|\eth w|^2_{lm}$, $\cle_{lm}$ and $\clb_{lm}$
as follows. First compute the spin-1 field, $\eth w$; this can be done, for
example, by synthesising $\sum_{lm} \sqrt{l(l+1)} w_{lm} {}_1 Y_{lm}(\vnhat)$
with fast spherical transforms (on an iso-latitude pixelisation).
Next, form the spin-zero field  $|\eth w|^2$ and the spin-2 field
$(\eth w)^2$ in real space by taking the squared modulus and the square
of $\eth w$ respectively. The scalar multipoles $|\eth w|^2_{lm}$ can then be
extracted from the spin-zero field with a (scalar) spherical transform,
while $\cle_{lm}$ and $\clb_{lm}$ are just the electric and magnetic
multipoles of the spin-2 field which can be extracted with spin-2 spherical
transforms. At each $l$ and $l'$, the $3j$ symbols for all the $L$ values
required can be computed efficiently with stable recursion
relations~\citep{schulten76}.

\subsection{Examples}
\label{subsec:ex2}

\begin{figure*}
\epsfig{figure=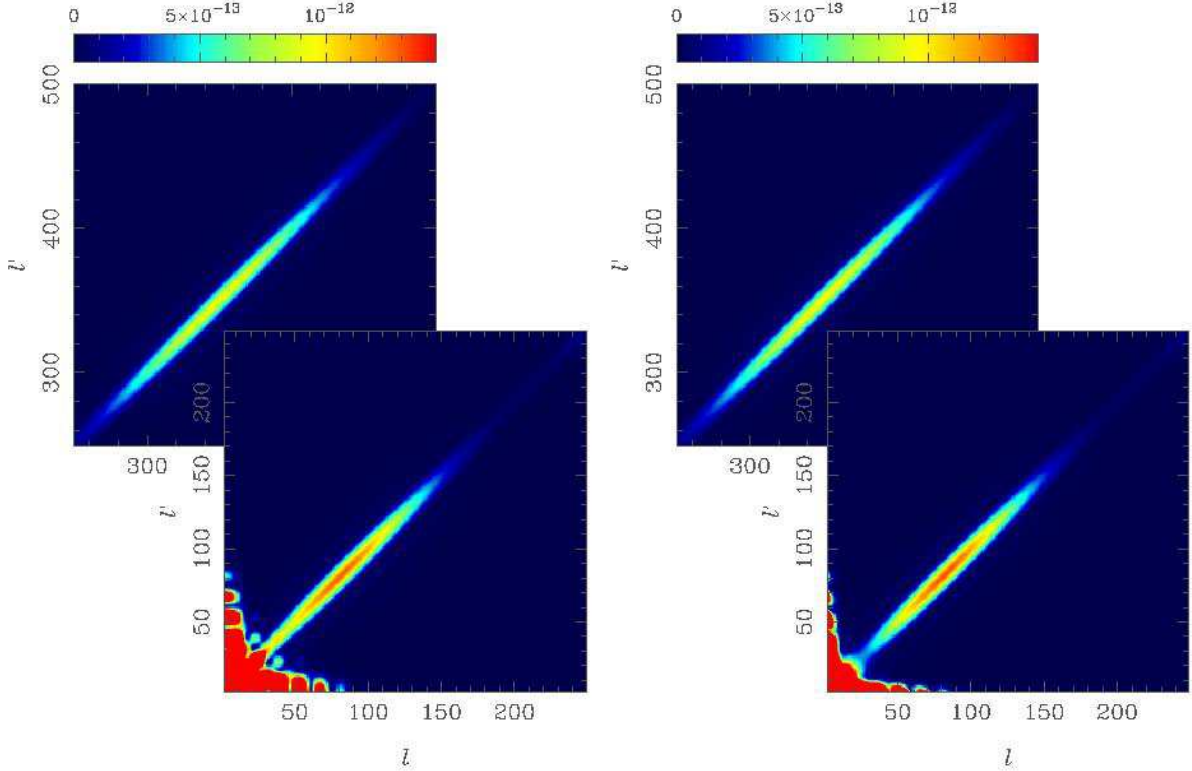,angle=0,width=16cm}
\caption{Blocks of the covariance matrix
$\rmn{cov}(\tilde{C}_l^E, \tilde{C}_{l'}^E)$ for
a $15\degr$ radius region with the same weighting as in
Fig.~\ref{fig:power_and_window}.
The exact covariance matrix is shown on the left, and its
approximation, equation~(\ref{eq:5-58}), on the right.
\label{fig:exact_and_approxEE_10_15_cosine}
}
\end{figure*} 

In this subsection we test the approximate covariance matrices developed
above against the exact matrices. Since computing the latter on intermediate
and small scales is prohibitively slow for general weight functions,
we restrict ourselves to the two azimuthally-symmetric examples considered in
Section~\ref{subsec:ex1}. Of course, for any practical application, our
approximate covariance matrices should be carefully verified against those
obtained from high-volume Monte-Carlo simulations with the chosen survey
geometry.

Figure~\ref{fig:exact_and_approxEE_10_15_cosine} shows the covariance
matrix $\rmn{cov}(\tilde{C}_l^E,\tilde{C}_{l'}^E)$ for the $15\degr$-radius
survey computed with the approximation~(\ref{eq:5-58}) and with the exact
expression, equation~(\ref{eq:5-37}).
Again, we take the weighting to be uniform
except for cosine-squared apodization of the outer $5\degr$ annulus.
The approximation is very accurate for the covariance of $\tilde{C}_l^E$,
except on the largest scales,
as expected given that our largest potential source of error -- 
the perturbative treatment of $E$-$B$ mixing -- is irrelevant in this case
since $C_l^E \gg C_l^B$. For $l>55$, the error on the diagonal is better
than 2\%\ except around the acoustic trough in
$C_l^E$, at $l \sim 200$, where it is around $-8$\%. Quite generally, the
error on the diagonal is largest at the acoustic peaks and troughs;
the approximation over-estimates at the peaks, where the curvature of
$C_l^E$ is negative, under-estimates at the troughs, and is best in between
where the curvature vanishes. This is consistent with the errors that we
would expect from removing $C_l^E$ from the convolution in
equation~(\ref{eq:5-37}). The magnitude of the correlations for
$\tilde{C}_l^E$ fall to lower than 1\%\ for $| l - l'| \ga 60$. For
$l$ and $l' > 55$, the maximum error in those off-diagonal elements with
correlations greater than 1\%\ is $\sim 30\%$, but the error is localised
to a few elements at the extremes of the diagonal band around the acoustic
trough at $l \sim 200$. This is, however, a very stringent error
criterion: to get a 1\%-correlated element accurate to 30\%\ with
Monte-Carlo methods would require $\sim 10^5$ simulations. A more relevant
measure for the off-diagonal elements is the absolute error in the
correlations, since large fractional errors in elements with small correlation
are relatively harmless. With this measure, the maximum error in the
correlations for $l$ and $l' > 55$ is $0.03$. We have made similar comparisons
for the case of the $\pm 20\degr$ Galactic cut described in
Section~\ref{subsec:ex1}. In this case, the error on the diagonal is
better than 1\%\ for $l \ge 55$. The only off-diagonal elements with
correlations $\ge 1\%$ are for $l'= l\pm 2$ and $l'=l\pm 4$, and for all
of these the fractional error is less than 2\%.

\begin{figure*}
\epsfig{figure=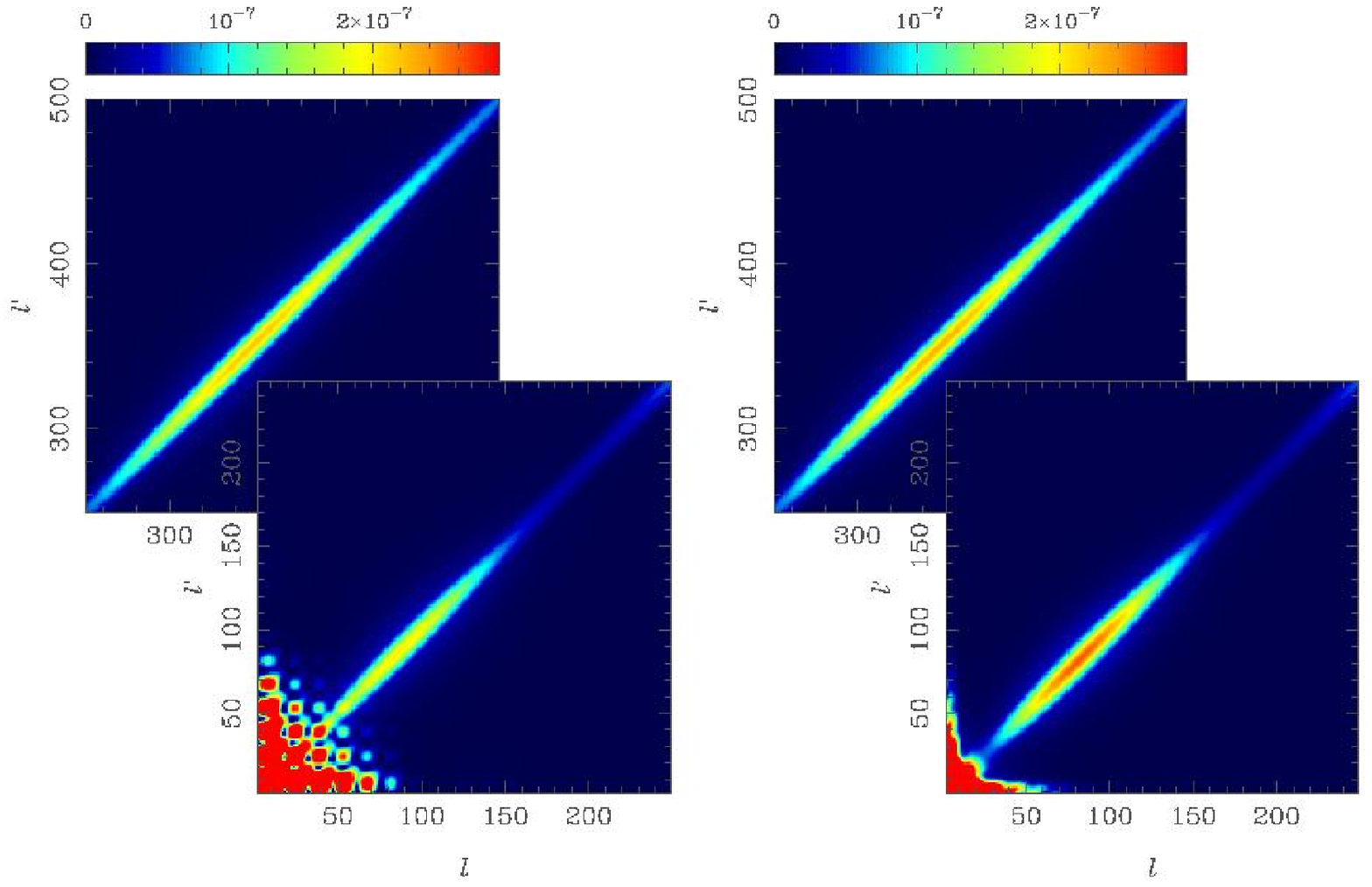,angle=0,width=16cm}
\caption{As Fig.~\ref{fig:exact_and_approxEE_10_15_cosine} but for
the covariance matrix
$l(l+1)l'(l'+1)\rmn{cov}(\tilde{C}_l^B, \tilde{C}_{l'}^B)$.
(Note the colour scale in this figure differs slightly from that in
Fig.~\ref{fig:exactBB_10_15_cosine_and_exactBB_10_15_cosine_Bonly}.)
\label{fig:exact_and_approxBB_10_15_cosine}
}
\end{figure*}

\begin{figure*}
\begin{center}
\epsfig{figure=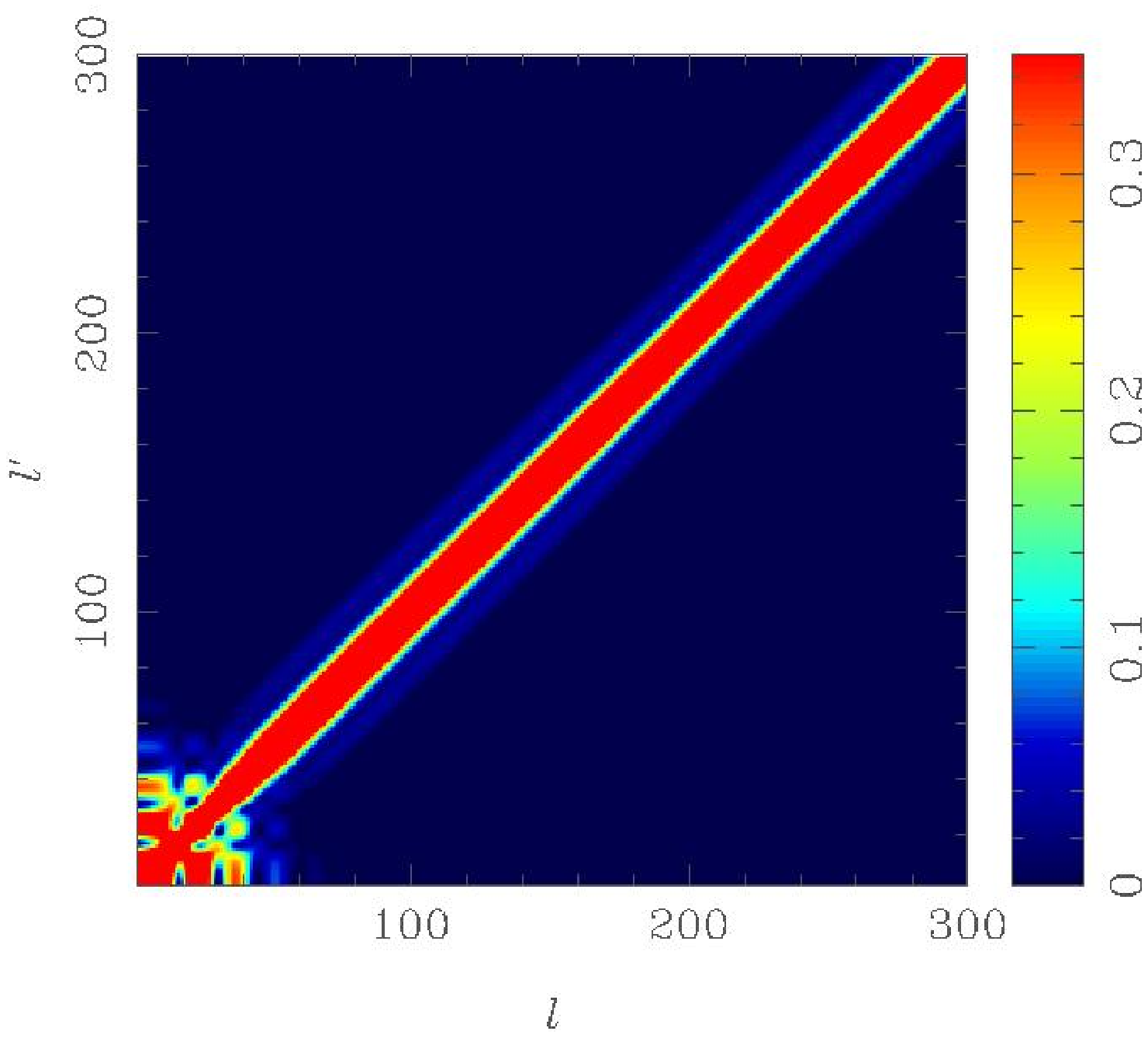,angle=0,width=5cm}
\epsfig{figure=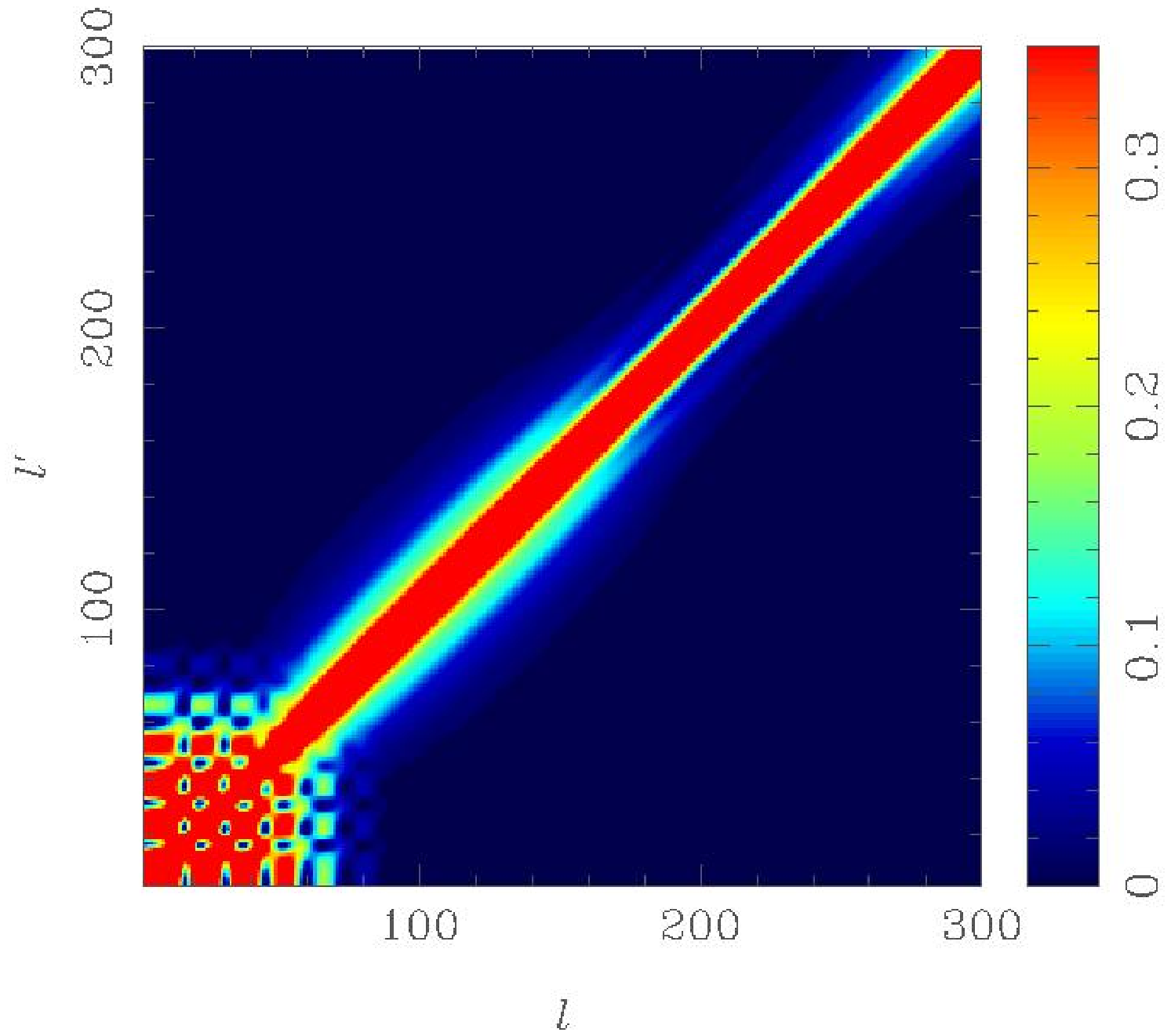,angle=0,width=5cm}
\epsfig{figure=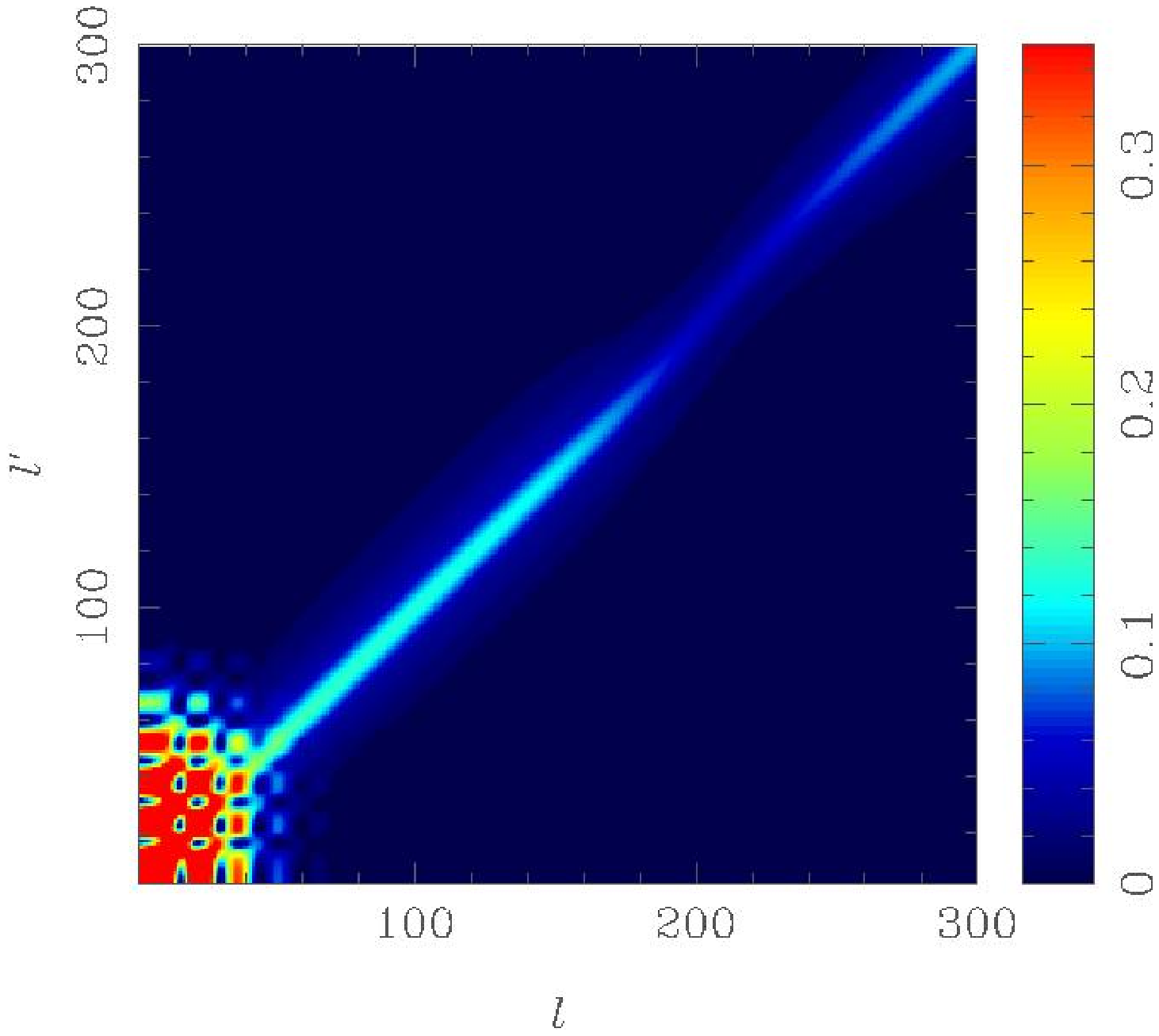,angle=0,width=5cm}
\end{center}
\caption{The exact correlation matrices
$\rmn{corr}(\tilde{C}_l^E,\tilde{C}_{l'}^E)$
(left), $\rmn{corr}(\tilde{C}_l^B,\tilde{C}_{l'}^B)$ (middle)
and $\rmn{corr}(\tilde{C}_l^E,\tilde{C}_{l'}^B)$ (right)
for the $15\degr$-radius survey. The colour scale has been chosen so that
it is saturated by elements with correlation $>0.35$ to emphasise the
broadening of the correlations for $\tilde{C}_l^B$ around the acoustic peaks
of $C_l^E$.
\label{fig:exact_corr_10_15_cosine}
}
\end{figure*}

The same comparisons are made for $\rmn{cov}(\tilde{C}_l^B,\tilde{C}_{l'}^B)$
for the $15\degr$-radius survey in
Fig.~\ref{fig:exact_and_approxBB_10_15_cosine}. The dominant contribution
to the covariance matrix over the full range plotted is now from
$E$-mode power that is mixed into $\tilde{B}_{lm}$; see
Fig.~\ref{fig:exactBB_10_15_cosine_and_exactBB_10_15_cosine_Bonly}. As
expected, the approximate covariance matrix for $\tilde{C}_l^B$ is
less accurate than for $\tilde{C}_l^E$ since any errors in our modelling
of the mixing are important for the former except at very large $l$. The
implied improvement in accuracy with increasing $l$ is apparent in the figure.
More quantitatively, for $l > 150$ the error on the diagonal is better
than 10\%. Interestingly, the error follows the same trend as for
$\tilde{C}_l^E$, i.e.\ it is largest in magnitude at the acoustic peaks
and troughs of $C_l^E$, and vanishes at the inflection points. This suggests
that the main source of error here is from removing $C_l^E$ from the
convolution in equation~(\ref{eq:5-38}) rather than errors in approximating
${}_- I^2(w)$ by equation~(\ref{eq:5-56}). (Since ${}_- I_{(lm)(lm)'}$ is
broader than ${}_+ I_{(lm)(lm)'}$, the fractional error made in
approximating the convolution is worse for the covariance of
$\tilde{C}_l^B$ than for $\tilde{C}_l^E$.) Regarding the off-diagonal
elements, for $l$ and $l' > 70$ the error \emph{in the correlation}
is $< 0.05$ for all elements. The correlation matrix is thus accurately
determined except on the largest scales, and so, in a practical application,
could be used as a template for the shape of the covariance matrix with
the variances calibrated off simulations. This procedure is similar to
that adopted by \emph{WMAP} for the analysis of the one-year temperature
data~\citep{hinshaw03}.
The $\Delta l = |l-l'|$ over
which correlations are significant is scale-dependent for the covariance
of $\tilde{C}_l^B$ unlike that for $\tilde{C}_l^E$; see
Fig.~\ref{fig:exact_corr_10_15_cosine}. The correlations
are broadest at the peaks of $C_l^E$, where the contribution of $E$ modes
to the $\tilde{C}_l^B$ covariance is greatest. This behaviour arises from
the different widths of the ${}_\pm I_{(lm)(lm)'}$ matrices, and would be
lost in any approximation that did not model the $E$-$B$ mixing accurately.
Repeating the comparison for the case of the Galactic cut, we find
that the diagonal elements are accurate to better than 10\%\ for
$l>50$, and better than 1\%\ for $l > 150$. For the off-diagonal elements,
the correlations are accurate to better than $0.03$ for $l$ and $l' > 55$.

\begin{figure*}
\epsfig{figure=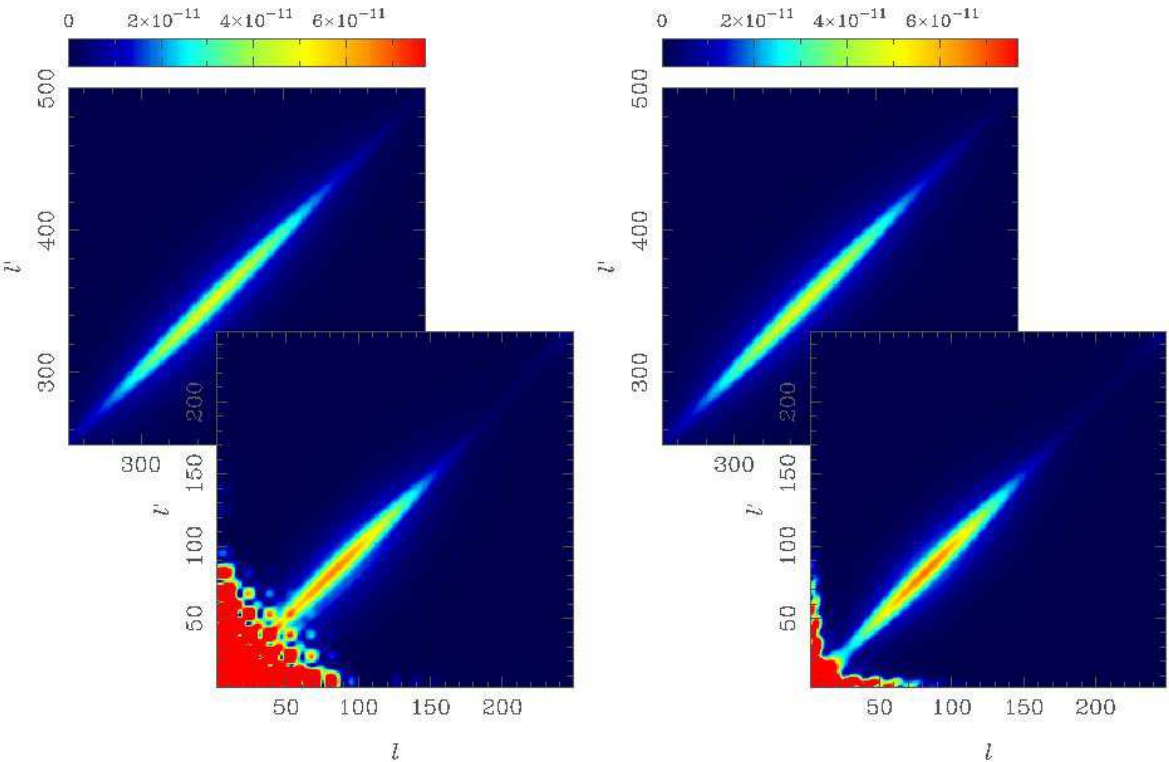,angle=0,width=16cm}
\caption{As Fig.~\ref{fig:exact_and_approxEE_10_15_cosine} but for
the covariance matrix
$ll'\rmn{cov}(\tilde{C}_l^E, \tilde{C}_{l'}^B)$.
\label{fig:exact_and_approxEB_10_15_cosine}
}
\end{figure*}

Finally, we compare the exact and approximate $\rmn{cov}(\tilde{C}_l^E,
\tilde{C}_{l'}^B)$ for the $15\degr$-radius survey in
Fig.~\ref{fig:exact_and_approxEB_10_15_cosine}. The error on the diagonal
is better than 10\%\ for all $l > 55$, and, as above, arises mainly from
removing $C_l^E$ (which dominates)
from the convolution in equation~(\ref{eq:5-39}), rather than
our ignoring the matrix commutator $[{}_+I(w),{}_-I(w)]$ in
equation~(\ref{eq:5-48}). For parameter estimation, the error in the
correlation matrix,
\begin{equation}
\rmn{corr}(\tilde{C}_l^E,\tilde{C}_{l'}^B) \equiv
\rmn{cov}(\tilde{C}_l^E,\tilde{C}_{l'}^B)/\sqrt{\rmn{var}(\tilde{C}_l^E)
\rmn{var}(\tilde{C}_{l'}^B)},
\end{equation}
is the more relevant quantity. The
exact correlation matrix is shown in
Fig.~\ref{fig:exact_corr_10_15_cosine}; there are strong correlations
on large scales due to the power in $C_l^E$ from reionization, but
for $l$ and $l' > 55$, the maximum correlation is only $0.15$ (and occurs on
the diagonal). Forming the approximate covariance matrix with
equations~(\ref{eq:5-58}), (\ref{eq:5-59}) and (\ref{eq:5-65}), we find that
the error is better than $0.03$ for all $l$ and $l' > 55$. For the case of
the Galactic cut, the correlations are $\la 0.01$ for $l$ and $l' > 55$, and
the maximum error in our approximate correlation matrix is $\sim 0.001$ over
the same range.

To summarise, the dominant error in our approximate covariance matrices,
for the weight functions tested here, is from the treatment of the power
spectra in the convolutions in equations~(\ref{eq:5-37})--(\ref{eq:5-39}).
The biggest errors on large and intermediate scales arise from the terms with
the dominant $C_l^E$ convolved with the broad ${}_- I_{(lm)(lm)'}$, and
so are most critical for the variance of $\tilde{C}_l^B$. In forming
the correlation matrices, the convolution error largely cancels and so
our approximate correlation matrices are more accurate than the covariances.
For small surveys it may be necessary to re-calibrate our approximate
variances off simulations. Accurate covariance matrices should then be
achievable by combining our approximate correlations matrices with the
calibrated variances.

\subsection{Approximations for band-power variances}
\label{subsec:rot}

In this subsection we use our approximate expressions for the covariances
of the pseudo-$C_l$s, equations~(\ref{eq:5-58}), (\ref{eq:5-59}) and
(\ref{eq:5-65}) to derive useful `rules of thumb' for the variances of
quasi-uncorrelated band-power estimates of the power spectra. Our results
generalise equations~(\ref{eq:flat50}) and (\ref{eq:flat51}), derived
earlier for Gaussian weighting in the flat-sky limit, and extend
the temperature result of~\citet{hivon02}, given here as
equation~(\ref{eq:hivon}), to polarization.

Following our treatment of Gaussian weighting in Section~\ref{subsec:gaussian},
we approximate the inversion from pseudo-$C_l$s to estimated $C_l$s by
the local transformation
\begin{equation}
\hat{C}_l^E = \frac{1}{N_l}(\alpha_l \tilde{C}_l^E - \beta_l \tilde{C}_l^B),
\quad 
\hat{C}_l^B = \frac{1}{N_l}(\alpha_l \tilde{C}_l^B - \beta_l \tilde{C}_l^E),
\label{eq:rot0}
\end{equation}
where we have defined $\alpha_l \equiv \sum_{ll'} P_{ll'}$ and
$\beta_l \equiv \sum_{l'} M_{ll'}$. Recall that e.g.\ $\sum_{ll'}P_{ll'}
C_{l'}^E$ is the contribution of $E$ modes to the mean of $\tilde{C}_l^E$,
and $\sum_{l'} M_{ll'} C_{l'}^B$ is the contribution from $B$ modes.
The normalisation $N_l \equiv \alpha_l^2 - \beta_l^2$. The covariance of
the $\hat{C}_l$s is thus a scaled version of that for the pseudo-$C_l$s.
As in the flat-sky limit, we
consider only the two extreme cases, $C_l^E=0$ and $C_l^B=0$, to keep the
algebra manageable. Such limits are sufficient to get the dominant
contribution to the variance of $\hat{C}_l^E$, and to assess the importance
of $E$-$B$ mixing for the variance of $\hat{C}_l^B$. We only need perform
the calculation for $C_l^B = 0$ since the other case then follows by symmetry.

We obtain quasi-uncorrelated estimates by averaging the recovered $C_l$s in
bands of width $\Delta l \ga l_{\rmn{max}}$, so we begin by averaging
the pseudo-$C_l$ covariances in such bands. Consider first the variance
of the binned $\tilde{C}_l^E$ for $l \gg l_\rmn{max}$.
The leading-order contribution is from the term involving
$(w^2)_{LM}$ in equation~(\ref{eq:5-58}).
To perform the summation over $l'$ in a
band centred on $\lbar$, we first insert a factor 
$(2l'+1)/(2\lbar+1)$, which is effectively unity for $l'$ in
the band provided
that $\lbar\gg \Delta l$. So long as we further choose bands that are
broad compared to $l_{\rmn{max}}$, we can reduce the lower limit of the
summation over $l'$ to $2$ and increase the upper limit to $\infty$.
(Recall that $l_{\rmn{max}}$ sets the range over which the pseudo-$C_l$s are
significantly correlated.) Finally, if we approximate $C_{l'}^E$ as constant
over a band we can replace it by $C_{\lbar}^E$, and the summation over the
square of the $3j$ symbol can be done in the same way as in the derivation of
equations~(\ref{eq:5-35a}) and (\ref{eq:5-35}). We find that, at leading-order,
\begin{equation}
\sum_{l'= \lbar-\Delta l/2}^{\lbar+\Delta_l/2} \rmn{cov}
(\tilde{C}_l^E,\tilde{C}_{l'}^E) \approx \frac{2}{2\lbar+1}
\frac{C_l^E C_{\lbar}^E}{4\pi} \sum_{LM} |(w^2)_{LM}|^2 = 
\frac{2}{2\lbar+1} w^{(4)}f_{\rmn{sky}} C_l^E C_{\lbar}^E.
\label{eq:rot1}
\end{equation}
To compute the variance of the binned $\tilde{C}_l^E$, we further average over
$l$ in the same band, treating $C_l^E$ as constant, to find
\begin{equation}
\rmn{var}(\tilde{C}_{\lbar}^E) \approx \frac{2}{(2\lbar+1)\Delta l}
w^{(4)}f_{\rmn{sky}} {C_{\lbar}^E}^2 \quad \mbox{($C_l^B=0$)}.
\label{eq:rot2}
\end{equation}
This result has the same form as that for the temperature anisotropies.

For the variance of the binned $\tilde{C}_{\lbar}^B$ we must retain all
terms in equation~(\ref{eq:5-59}) that survive setting $C_l^B=0$. We perform
the summation over $l'$ in a similar manner to that described above, making
use of some additional results for summing products of $3j$ symbols that
are given in Appendix~\ref{app:useful_sums}. Further averaging $l$ within
the band, we find, at leading order,
\begin{equation}
\rmn{var}(\tilde{C}_{\lbar}^B) \approx \frac{1}{(2\lbar+1)\lbar^2
(\lbar+1)^2 \Delta l} \frac{{C_{\lbar}^E}^2}{\pi}
\sum_{LM} \left(
2 \left| |\eth w|^2_{LM} \right|^2 + |\cle_{LM}|^2 + |\clb_{LM}|^2 \right)
\quad \mbox{($C_l^B=0$)}.
\label{eq:rot3}
\end{equation}
This can be simplified further by noting that
\begin{equation}
\sum_{LM} \left| |\eth w|^2_{LM} \right|^2
 = \int |\eth w|^4 \, \ud \vnhat = \int (\nabla w)^4 \, \ud \vnhat,
\label{eq:rot4}
\end{equation}
and, from equation~(\ref{eq:5-60}) and the completeness of the spin-weight
harmonics,
\begin{equation}
\sum_{LM} \left( |\cle_{LM}|^2 + |\clb_{LM}|^2 \right) = \frac{1}{2}
\sum_{LM} \left[ |{}_2(\eth w)^2_{LM}|^2 +|{}_{-2}(\mybeth w)^2_{LM}|^2 \right]
= \int |\eth w|^4 \, \ud \vnhat. 
\label{eq:rot5}
\end{equation}
Putting these results together, we find
\begin{equation}
\rmn{var}(\tilde{C}_{\lbar}^B) \approx \frac{3}{(2\lbar+1)\lbar^2
(\lbar+1)^2 \Delta l} \frac{{C_{\lbar}^E}^2}{\pi}
\int (\nabla w)^4 \, \ud \vnhat \quad \mbox{($C_l^B = 0$)}.
\label{eq:rot6}
\end{equation}
Note that in the limit $C_l^B=0$, the variance of $\tilde{C}_l^B$
arises solely from $E$-$B$ mixing which explains why only
derivatives of $w$ appear in this equation.

The final term we require is $\rmn{cov}(\tilde{C}_{\lbar}^E,
\tilde{C}_{\lbar}^B)$. Setting $C_l^B=0$ in equation~(\ref{eq:5-65})
and using similar approximations to those adopted above, the leading-order
result is
\begin{eqnarray}
\rmn{cov}(\tilde{C}_{\lbar}^E, \tilde{C}_{\lbar}^B) &\approx&
\frac{1}{(2\lbar+1)\lbar(\lbar+1)\Delta l}\frac{{C_{\lbar}^E}^2}{%
4\pi} \sum_{LM} L(L+1)|(w^2)_{LM}|^2 \nonumber\\
&=& \frac{1}{(2\lbar+1)\lbar(\lbar+1)\Delta l}\frac{{C_{\lbar}^E}^2}{%
4\pi} \int (\nabla w^2)^2 \, \ud\vnhat. 
\label{eq:rot7}
\end{eqnarray}
As expected, the relative sizes of $\rmn{var}(\tilde{C}_{\lbar}^E)$,
$\rmn{cov}(\tilde{C}_{\lbar}^E,\tilde{C}_{\lbar}^B)$ and
$\rmn{var}(\tilde{C}_{\lbar}^B)$ for $C_l^B=0$ are roughly
$1$, $O(l_{\rmn{max}}^2/ \lbar^2)$ and $O(l_{\rmn{max}}^4/\lbar^4)$
respectively.

We now have all the results required to compute the leading-order variances
of the band-power estimates for $C_l^E$ and $C_l^B$. Combining
the (co)variances calculated above with equation~(\ref{eq:rot0}), we
find
\begin{eqnarray}
\rmn{var}(\hat{C}_{\lbar}^E) &\approx& \frac{2 w^{(4)} {C_{\lbar}^E}^2}{%
(2\lbar+1)\Delta l f_{\rmn{sky}} {w^{(2)}}^2} \quad \mbox{($C_l^B=0$)} ,
\label{eq:rot8} \\
\rmn{var}(\hat{C}_{\lbar}^B) &\approx& \frac{2 {C_{\lbar}^E}^2}{%
(2\lbar+1) \Delta l f_{\rmn{sky}}} \frac{6}{\lbar^2(\lbar+1)^2}
\left( \frac{(\nabla w)^{(4)}}{{w^{(2)}}^2} + \frac{2}{3}
\frac{w^{(4)} {(\nabla w)^{(2)}}^2}{{w^{(2)}}^4}
- \frac{4}{3} \frac{(\nabla w)^{(2)} (w\nabla w)^{(2)}}{{w^{(2)}}^3} \right)
\quad \mbox{($C_l^B=0$)}.
\label{eq:rot9}
\end{eqnarray}
Here, we have introduced the convenient notation $4\pi X^{(i)} f_{\rmn{sky}}
\equiv \int X^i \, \ud \vnhat$, where $X=w$, $\nabla w$ or $w\nabla w$ in
the above, and approximated $\alpha_l \approx w^{(2)} f_{\rmn{sky}}$ and
$\beta_l \approx 2 (\nabla w)^{(2)} f_{\rmn{sky}} / [l (l+1)]$ [see
equations~(\ref{eq:5-35a}) and (\ref{eq:15})].

To obtain the equivalent results for $C_l^E=0$, one should swap
$E$ and $B$ everywhere in equations~(\ref{eq:rot8}) and (\ref{eq:rot9}).
It is straightforward to verify that for a Gaussian weight we recover the
variances given in Section~\ref{subsec:gaussian} in the flat-sky limit.
We now see, quite generally, that the ratio of the excess sample variance
from $E$ modes in band-power estimates of $C_l^B$ to the variance from
$B$ modes is $\sim (C_l^E/C_l^B)^2 O(l_\rmn{max}^4/l^4)$. We consider the
implications of this for $B$-mode detection with pseudo-$C_l$ techniques
in the next section.

\section{Implications for $B$-mode detection}
\label{sec:implications}

The realisation that, in linear theory, scalar perturbations do not
generate $B$-mode polarization~\citep{kamionkowski97,seljak97} has
paved the way for a future generation of instruments designed for
high-sensitivity searches for gravitational waves via
$B$-mode polarization. Instrument noise
can be reduced to the $\sim 4\,\mu\rmn{K}$-arcmin level by surveying e.g.\
a few-hundred square degrees for around a year with arrays of hundreds of
detectors. With instrument noise at this level, the sample variance from the
$B$ modes produced by gravitational lensing of the scalar $E$ modes dominates
the thermal noise in the random error budget for $C_l^B$, assuming perfect
isolation of $B$ modes. (We are assuming the tensor-to-scalar ratio
$r \ll 0.01$ here, so that sample variance of the linear $B$ modes is
sub-dominant.) If we make no attempt to reconstruct and subtract the
lensing signal, its variance sets a fundamental limit to the
one-sigma error of $\Delta r=1.4\times 10^{-5}$ for a full-sky survey in
the null hypothesis, $r=0$. This should be compared with the limit
of only $0.02$ obtainable by analysing temperature and electric polarization
only. The limit on $r$ is much poorer in this case because of the large
sample variance of the dominant scalar perturbations.

\begin{figure*}
\epsfig{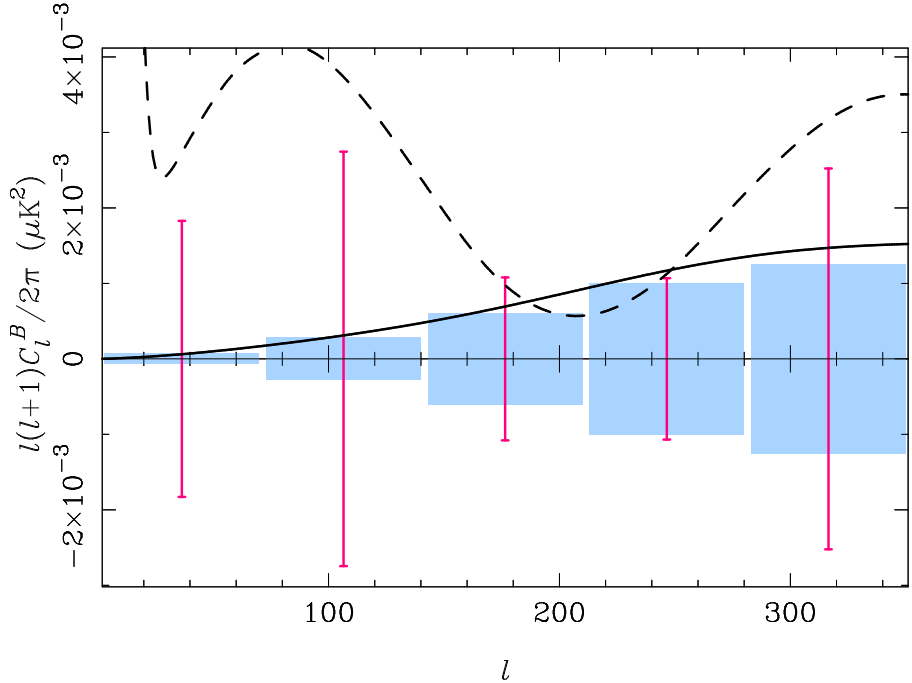}
\caption{Sample variance errors on the recovered $\hat{C}_l^B$ using the
estimator of~\citet{chon04}; see also Section~\ref{sec:inverse}.
The survey area and weight function is the same as in 
Fig.~\ref{fig:power_and_window}.
The error boxes are the contribution purely from $C_l^B$ to the
one-sigma errors on flat band-power estimates with a $\Delta l = 70$. These
are thus representative of the errors that would be obtained if $E$ and $B$
modes were separated (without loss) at the level of the map. They agree well
with the simple rule of thumb in equation~(\ref{eq:rot8}), plotted as the
solid line.
The error bars are the contribution purely from $C_l^E$. They agree
reasonably with the dashed line, which is the rule of thumb in
equation~(\ref{eq:rot9}). Critically, these
dominate the errors due to $C_l^B$ in the angular range relevant to
gravity-wave searches with $B$-mode polarization.
All errors are computed in the null hypothesis $r=0$, so the contribution from
$C_l^B$ arises only from sample variance of the lens-induced $B$ modes.
\label{fig:errors_on_recovered_B}
}
\end{figure*}

If pseudo-$C_l$ techniques are to be applied to future,
high-sensitivity searches for gravitational waves, it is clearly desirable
that the imperfect isolation of $B$ modes inherent in such methods
should not compromise sensitivity to $r$. In
Fig.~\ref{fig:errors_on_recovered_B} we plot the errors on
flat band-power estimates of $C_l^B$ (with $\Delta l = 70$)
obtained by using the
unbiased estimator of~\citet{chon04}; see also the discussion in
Section~\ref{sec:inverse}. The weight function applied to the
$15\degr$-radius survey, and apodization of the correlation function,
are the same as in Fig.~\ref{fig:power_and_window}. The errors are obtained
from the exact covariance matrices for the pseudo-$C_l$s by transforming
(two-sidedly) with the appropriate pseudo-inverses $\tilde{P}^{-1}_{ll'}$
and $\tilde{M}^{-1}_{ll'}$. These pseudo-inverse matrices are shown in
Fig.~\ref{fig:power_and_window}. The errors are calculated for $r=0$ in two
ways. First, we set $C_l^E=0$ so that we obtain the error in the recovered
$C_l^B$ due to sample variance of the lensing $B$ modes. These errors
are representative of what would be achieved if we were able to separate $E$
and $B$ modes perfectly, and without loss, on all scales in the maps;
they agree very well with
the rule of thumb in equation~(\ref{eq:rot8}) if we swap the roles of $E$ and
$B$ there. We also calculate the band-power errors obtained by
setting $C_l^B=0$, i.e.\ removing the $B$-mode power from lensing; they
agree reasonably with the rule of thumb in equation~(\ref{eq:rot9}). These
errors arise because, in any realisation, our chosen estimator $\hat{C}_l^B$
receives a contribution from both $E$ and $B$ modes, despite
only coupling to $B$-mode power in the mean. Such errors can only be
removed by isolating $B$ modes at the level of the map,
or by adopting more optimal, non-local weighting of the data, as in
maximum-likelihood power spectrum estimation. Critically, we see from
Fig.~\ref{fig:errors_on_recovered_B} that the errors due to the sample variance
of the $E$ modes that leak into the estimator dominate the sample variance
from lensing $B$ modes on large scales, where the gravitational-wave signal
resides. We conclude from the figure that pseudo-$C_l$ estimators will be
highly non-optimal for future experiments with thermal sensitivity better than
$\sim 4\, \mu\rmn{K}$-arcmin, targeting $B$-mode polarization over a
few-hundred square degrees.\footnote{%
If gravitational lensing were absent, the optimal survey for detecting
$B$-mode polarization with a power spectrum of known shape but (low) unknown
amplitude is a circle of radius around 10 degrees~\citep{jaffe00,lewis02}.
This assumes that the sample variance of the $B$ modes from gravitational
waves does not dominate the error budget.
If the lens-induced $B$ modes are treated as
an additional (Gaussian) noise term, the optimal survey area is increased
since sample variance becomes an issue again.
The best compromise is roughly to balance the sample variance of
the lens-induced $B$ modes with the thermal-noise variance. As a specific
example, for 200 \emph{pairs} of polarization-sensitive bolometers, with each
bolometer having sensitivity $200\, \mu \mathrm{K}\sqrt{\mathrm{s}}$,
and one year of integration, a survey covering two per cent of the sky
balances the thermal and sample variance.}

We can make a quantitative estimate of the impact of the additional sample
variance due to leaked  $E$ modes on the detectability of gravitational
waves as follows. To preserve all information in the pseudo-$C_l$s,
we bypass the step of forming unbiased estimates of $C_l^B$
[which involves the singular transformation in equation~(\ref{eq:estimates})],
and instead estimate $r$ directly from the pseudo-$C_l$s. We
concatenate $\tilde{C}_l^E$ and $\tilde{C}_l^B$ into a single data
vector $\tilde{C}_l$, whose covariance matrix
$\rmn{cov}(\tilde{C}_l,\tilde{C}_{l'})$ then consists of blocks formed from
$\rmn{cov}(\tilde{C}_l^E,\tilde{C}_{l'}^E)$,
$\rmn{cov}(\tilde{C}_l^B,\tilde{C}_{l'}^B)$ and
$\rmn{cov}(\tilde{C}_l^E,\tilde{C}_{l'}^B)$. We assume that all
parameters except $r$ are fixed, and calculate its one-sigma
Fisher error from
\begin{equation}
\frac{1}{(\Delta r)^2} = \sum_{ll'} \frac{\partial\langle \tilde{C}_l \rangle}
{\partial r} \frac{\partial\langle \tilde{C}_{l'} \rangle}
{\partial r} \rmn{cov}^{-1}(\tilde{C}_l, \tilde{C}_{l'}). 
\end{equation}
In practice, we invert the covariance matrix with a singular-value
decomposition which projects out any linear dependencies in the
$\tilde{C}_l$ due to the finite sky coverage. For the $15\degr$-radius
survey, the one-sigma error on $r$ in the null hypothesis ($r=0$) is
$\Delta r = 0.053$. Our sensitivity to $r$ improves dramatically to
$\Delta r = 1.096\times 10^{-4}$ if we artificially set $C_l^E=0$
(and $\partial C_l^E / \partial r = 0$) in the analysis.
This latter case is representative of what could be achieved with
lossless separation of the $B$ modes, followed by a pseudo-$C_l$ analysis
of the Stokes maps of these $B$ modes. In practice, $E$-$B$ separation
is lossy, particularly on the largest scales where there is significant
$B$-mode power from reionization~\citep{lewis02,bunn03,lewis03}.
If we assume, crudely, that
large-scale modes with $l<20$ are lost to the separation, we find
a $\Delta r \sim 2\times 10^{-3}$ for the $15\degr$-radius survey. The
effect of not isolating the $B$ modes properly in the
pseudo-$C_l$ analysis thus degrades the limit on $r$ by a factor
$\sim 25$.

\section{Conclusion}
\label{sec:conclusion}

Accurate calculation of the covariance matrix is an important part of
power spectrum estimation during the analysis of CMB data. A `correct'
estimator (i.e.\ unbiased), but with poorly quantified errors, is useless
for subsequent parameter estimation. In this paper we have constructed
the contribution from sample variance to the covariance matrices
of heuristically-weighted quadratic estimates of the CMB polarization
power spectra. Exact calculation of these matrices for general weighting
is numerically feasible at large scales ($l$ less than a hundred or so),
but the $O(l^6)$ scaling is prohibitive on small scales. To resolve this
problem we developed analytic approximations to the polarization covariance
matrices, including the effects of mode coupling and, in a perturbative
manner, the mixing of $E$ and $B$ modes that inevitably arises with heuristic
weighting. Our approximations, which generalise earlier work on the
temperature anisotropies~\citep{hinshaw03,efstathiou04}, can be computed
efficiently to high $l$. Accurate covariance matrices can be obtained on
all scales by replacing the low-$l$ sectors of our approximations
with a numerical calculation.

We tested the approximations against numerical
results for two toy-model surveys for which azimuthal symmetry allows
efficient evaluation of the exact covariances. For the more challenging
survey, a $15\degr$-radius survey with cosine-squared apodization of the
outer $5\degr$, the most significant errors were on the variance of
$\tilde{C}_l^B$. These errors arise from our local treatment of the
$E$-mode power spectrum when it appears in a convolution with the broad kernel
that describes the mixing of $E$ and $B$ modes in equation~(\ref{eq:5-38}).
For the variance of
$\tilde{C}_l^E$ (and the temperature anisotropies) the approximation is
more accurate because the dominant convolution involves a kernel that is
intrinsically more narrow. It will probably
be very difficult to improve the error on the variance of $\tilde{C}_l^B$
in an analytic treatment. Fortunately, convolution errors tend to cancel
in the approximation of the \emph{correlation} matrices. We observed maximum
errors $\la 0.05$ in the correlations, except on the largest scales,
suggesting that, in a practical
application, it should be possible to construct accurate covariance matrices
by combining the approximate correlations with variances calibrated off
simulations.
By binning our approximate covariances, we also derived
useful rules of thumb for the variances of band-power estimates of
$C_l^E$ and $C_l^B$. These are particularly simple to evaluate, involving only
integrals of products of the weight function and its derivatives.

A further aim of this paper was to quantify the conditions under which
pseudo-$C_l$ methods can be applied to gravitational-wave
searches via $B$-mode polarization. Measurements of the $B$-mode
power spectrum from gravitational waves with future
surveys with sub-$4\,\mu\rmn{K}$-arcmin thermal noise are potentially
limited by the sample variance of
lens-induced $B$ modes. We showed that for surveys covering one or two per cent
of the sky, the additional variance due to leaked $E$ modes dominates the
lensing noise. Adopting pseudo-$C_l$ methods would therefore by highly
sub-optimal: the error on the tensor-to-scalar ratio would be $\sim 25$ times
higher than if $B$ modes had been isolated before forming quadratic estimates.
Ultimately, this additional variance would limit the tensor-to-scalar ratio
that is detectable with pseudo-$C_l$ methods to $\sim 0.05$. Map-level
$E$-$B$ separation, or more optimal weighting such as in maximum-likelihood
power spectrum estimation, will therefore be essential to ensure
the maximum returns from future $B$-mode surveys.

\section*{Acknowledgments}

ADC acknowledges a Royal Society University Research Fellowship.
GC acknowledges travel support from Queens' College, Cambridge and
the Cavendish Laboratory. All theoretical power spectra in this paper
were computed with CAMB~\citep{lewis00}.
Some of the 
results in this paper have been derived using the
LAPACK\footnote{http://www.netlib.org/lapack/} package for
matrix decompositions.

\appendix
\section{Correlators for Gaussian-weighted
$\tilde{E}(\vl)$ and $\tilde{B}(\vl)$}
\label{app:flat}

In this appendix we derive approximate analytic expressions for the
correlations between the flat-sky Fourier transforms
$\tilde{E}(\vl)$ and $\tilde{B}(\vl)$ for a Gaussian weight function
$w(\vx)$. Our expressions are valid for $l$ and $l'$ large compared
to $1/\sigma$. In this limit, the power spectra can be
approximated as being constant over the support of $w(\vl-\vL)$; see
Section~\ref{sec:approx-cov}. Using equations~(\ref{eq:flat6}) and
(\ref{eq:flat7}) we can approximate the correlators by
\begin{eqnarray}
\langle \tilde{E}(\vl) \tilde{E}^\ast(\vl') \rangle &\approx&
\frac{\sqrt{C_l^E C_{l'}^E}}{2\pi} \int \frac{d^2\vL}{2\pi}\,
{}_+I(\vl,\vL) {}_+I^\ast(\vl',\vL) + \frac{\sqrt{C_l^B C_{l'}^B}}{2\pi}
\int \frac{d^2\vL}{2\pi}\, {}_-I(\vl,\vL) {}_-I^\ast(\vl',\vL),
\label{eq:flat18} \\
\langle \tilde{B}(\vl) \tilde{B}^\ast(\vl') \rangle &\approx&
\frac{\sqrt{C_l^B C_{l'}^B}}{2\pi} \int \frac{d^2\vL}{2\pi}\,
{}_+I(\vl,\vL) {}_+I^\ast(\vl',\vL) + \frac{\sqrt{C_l^E C_{l'}^E}}{2\pi}
\int \frac{d^2\vL}{2\pi}\, {}_-I(\vl,\vL) {}_-I^\ast(\vl',\vL).
\label{eq:flat19}
\end{eqnarray}
For a Gaussian weight function we can evaluate the integrals in these
equations analytically; we find
\begin{eqnarray}
\int \frac{d^2\vL}{2\pi}\, {}_\pm I(\vl,\vL) {}_\pm I^\ast(\vl',\vL) &=&
\frac{1}{4}\sigma^2 \cos2(\phi_\vl-\phi_{\vl'})e^{-|\vl-\vl'|^2\sigma^2/4}
\pm \frac{1}{2}\sigma^2 \cos2(\phi_\vl + \phi_{\vl'} - 2 \phi_{\vl + \vl'})
\nonumber \\
&&\mbox{} \times
\left[\frac{-4 e^{-(l^2+l^{\prime 2})\sigma^2/2}}{|\vl+\vl'|^2\sigma^2}
\left( 1 + \frac{12}{|\vl+\vl'|^2 \sigma^2}\right)
+ e^{-|\vl-\vl'|^2\sigma^2/4} \left(\frac{1}{2}-\frac{8}{|\vl+\vl'|^2\sigma^2} 
+ \frac{48}{|\vl+\vl'|^4 \sigma^4} \right)\right].
\label{eq:flat20}
\end{eqnarray}
Note that
\begin{equation}
\int \frac{d^2\vL}{2\pi}\, [{}_+I(\vl,\vL) {}_+I^\ast(\vl',\vL) +
{}_-I(\vl,\vL) {}_-I^\ast(\vl',\vL)] = \frac{1}{2} \sigma^2
\cos2(\phi_\vl-\phi_{\vl'})e^{-|\vl-\vl'|^2\sigma^2/4},
\label{eq:flat21}
\end{equation}
and the right-hand side is ${}_+I(\vl,\vl')$ constructed from
$w^2(\vx)$. This result follows quite generally from equation~(\ref{eq:flat8}).
To calculate the covariance of the errors of the power spectrum estimates
we also need the following correlator
\begin{equation}
\langle \tilde{E}(\vl) \tilde{B}^\ast(\vl') \rangle \approx
i\frac{\sqrt{C_l^E C_{l'}^E}}{2\pi} \int \frac{d^2\vL}{2\pi}\,
{}_+I(\vl,\vL) {}_-I^\ast(\vl',\vL) + i\frac{\sqrt{C_l^B C_{l'}^B}}{2\pi}
\int \frac{d^2\vL}{2\pi}\, {}_-I(\vl,\vL) {}_+I^\ast(\vl',\vL).
\label{eq:flat22}
\end{equation}
The integrals in this expression evaluate to
\begin{eqnarray}
i \int \frac{d^2\vL}{2\pi}\, {}_\pm I(\vl,\vL) {}_\mp I^\ast(\vl',\vL) &=&
\frac{1}{4}\sigma^2 \sin2(\phi_\vl-\phi_{\vl'})e^{-|\vl-\vl'|^2\sigma^2/4}
\mp \frac{1}{2}\sigma^2 \sin2(\phi_\vl + \phi_{\vl'} - 2 \phi_{\vl + \vl'})
\nonumber \\
&&\mbox{} \times
\left[\frac{-4 e^{-(l^2+l^{\prime 2})\sigma^2/2}}{|\vl+\vl'|^2\sigma^2}
\left( 1 + \frac{12}{|\vl+\vl'|^2 \sigma^2}\right) 
+ e^{-|\vl-\vl'|^2\sigma^2/4} \left(\frac{1}{2}-\frac{8}{|\vl+\vl'|^2\sigma^2} 
+ \frac{48}{|\vl+\vl'|^4 \sigma^4} \right)\right].
\label{eq:flat23}
\end{eqnarray}
We now have
\begin{equation}
\int \frac{d^2\vL}{2\pi}\, [{}_+I(\vl,\vL) {}_-I^\ast(\vl',\vL) +
{}_-I(\vl,\vL) {}_+I^\ast(\vl',\vL)] = - \frac{i}{2} \sigma^2
\sin2(\phi_\vl-\phi_{\vl'})e^{-|\vl-\vl'|^2\sigma^2/4},
\label{eq:flat24}
\end{equation}
and the right-hand side is ${}_-I(\vl,\vl')$ constructed from
$w^2(\vx)$. Note that $\langle \tilde{E}(\vl) \tilde{B}^\ast(\vl') \rangle=0$
if $\vl$ and $\vl'$ are parallel or perpendicular.

\section{Useful sums of products of {$3j$} symbols}
\label{app:useful_sums}

In Sections~\ref{sec:properties} and~\ref{subsec:rot} we make use of a number
of results for summing products of $3j$ symbols. Some of these follow
trivially from the orthogonality properties of the $3j$s, but others require
more effort. Here, we sketch our derivation of these
latter results. We make extensive use of well-known properties of the $3j$
symbols, such as their symmetries and integral representations
(e.g.~\citealt{varshalovich}).

In Section~\ref{sec:properties} we require
\begin{equation}
S_1 \equiv \sum_{l'} (-1)^K (2l'+1) \left( \begin{array}{ccc} l & l' & L \\
                                                       -2 & 2 & 0
                                     \end{array} \right)^2.
\label{eq:appb1}
\end{equation}
It is the presence of the factor of $(-1)^K$, where $K\equiv l+l'+L$, that
complicates the evaluation of this summation. To proceed, we express
the product of $3j$ symbols as an integral of reduced Wigner matrices
using
\begin{equation}
\int_{-1}^{1}
d^{l_1}_{m_1 n_1}(\beta) d^{l_2}_{m_2 n_2}(\beta) d^{l_3}_{m_3 n_3}(\beta)
\, \ud \cos\beta = 2 \left( \begin{array}{ccc} l_1 & l_2 & l_3 \\
					       m_1 & m_2 & m_3 \end{array}
\right) \left( \begin{array}{ccc} l_1 & l_2 & l_3 \\
					       n_1 & n_2 & n_3 \end{array}
\right) \quad \mbox{(for $\sum_i m_i = \sum_i n_i = 0$ )},
\label{eq:appb1b}
\end{equation}
so that
\begin{equation}
S_1 = \frac{1}{2}\sum_{l'}(2l'+1) \int d^l_{-2 2}(\beta) d^{l'}_{2\, -2}(\beta)
d^L_{00}(\beta)\, \ud \cos\beta .
\label{eq:appb2}
\end{equation}
The factor $(-1)^K$ has been absorbed in reducing the product of
$3j$ symbols to a square.
The summation over $l'$ can be performed by first expressing $d^{l'}_{2\, -2}$
in terms of $d^{l'}_{22}$ using the recursion relations for the reduced
Wigner matrices~\citep{varshalovich}, and then using the completeness
relation. Full details can be found in~\citet{chon04}, but the final result
is
\begin{equation}
\half \sum_{l'} (2l'+1)d^{l'}_{2 -2}(\beta) = \delta(\cos\beta -1) +
\csc^2(\beta/2).
\label{eq:appb3}
\end{equation}
Inserting this into equation~(\ref{eq:appb2}) gives
\begin{equation}
S_1 = \int\csc^2(\beta/2) d^{l}_{2 -2}(\beta) d^L_{00}(\beta) \, \ud
\cos\beta. 
\label{eq:appb4}
\end{equation}
Using the differential representation of the reduced Wigner matrices
(see Section 4.3.2 of~\citealt{varshalovich}), it can be shown that
$\csc^2(\beta/2) d^l_{2 -2}(\beta)$ is a polynomial in $\cos\beta$ of degree
$l-1$. It can thus be expanded as the series
$\sum_{n=0}^{l-1} a_n d^n_{00}(\beta)$, i.e.\ an expansion in Legendre
polynomials. Using the orthogonality of the $d^l_{00}$, we see that
the integral in equation~(\ref{eq:appb4}) vanishes if $L \ge l$. It
can be performed for $L < l$ by replacing $d^l_{2 -2}$ by its
differential representation and repeatedly integrating by parts. Only
boundary terms then survive, and evaluating these establishes the final
result~\citep{chon04}:
\begin{equation}
\sum_{l'} (-1)^K (2l'+1) \left( \begin{array}{ccc} l & l' & L \\
                                                       -2 & 2 & 0
                                     \end{array} \right)^2 =
\left\{ \begin{array}{ll}
1- 4 \frac{L(L+1)}{l(l+1)} + 3 \frac{(L+2)!}{(L-2)!}\frac{(l-2)!}{(l+2)!} \quad
& \mbox{for $L \le l$} \\
0 & \mbox{otherwise.} \end{array} \right.
\label{eq:appb5}
\end{equation}

In Section~\ref{subsec:rot}, we require the evaluation of
\begin{equation}
S_2 \equiv \sum_{l'} (-1)^K (2l'+1) \left( \begin{array}{ccc} l & l' & L \\
                                                       -1 & 1 & 0
                                     \end{array} \right)^2.
\label{eq:appb6}
\end{equation}
This follows along similar lines to $S_1$ above. A useful intermediate result 
is
\begin{equation}
d^l_{1 -1}(\beta) = - d^l_{11}(\beta) + \csc^2(\beta/2) \int_0^\beta
\tan(\beta'/2) d^l_{11}(\beta')\, \ud \beta',
\label{eq:appb7}
\end{equation}
from which it follows that
\begin{equation}
\half \sum_{l'} (2l'+1) d^{l'}_{1 -1}(\beta) = - \delta(\cos\beta -1)
+ \half \csc^2(\beta/2).
\label{eq:appb8} 
\end{equation}
The summation $S_2$ thus reduces to
\begin{equation}
S_2 = \frac{1}{2} \int\csc^2(\beta/2) d^{l}_{1 -1}(\beta) d^L_{00}(\beta) \,
\ud \cos\beta,
\label{eq:appb9}
\end{equation}
which, again, vanishes if $L \ge l$. Inserting the differential representation
of $d^l_{1-1}$ and integrating by parts, we find

\begin{equation}
\sum_{l'} (-1)^K (2l'+1) \left( \begin{array}{ccc} l & l' & L \\
                                                       -1 & 1 & 0
                                     \end{array} \right)^2 =
\left\{ \begin{array}{ll}
1- \frac{L(L+1)}{l(l+1)} \quad
& \mbox{for $L \le l$} \\
0 & \mbox{otherwise.} \end{array} \right.
\label{eq:appb10}
\end{equation}
A further result we require in Section~\ref{subsec:rot} is
\begin{equation}
S_3 \equiv \sum_{l'} (-1)^K (2l'+1) \left( \begin{array}{ccc} l & l' & L \\
                                                       -1 & -1 & 2
                                     \end{array} \right)^2.
\label{eq:appb11}
\end{equation}
Using equations~(\ref{eq:appb1b}) and (\ref{eq:appb8}), we find
\begin{equation}
S_3 = \frac{1}{2}\int\csc^2(\beta/2) d^{l}_{1 -1}(\beta) d^L_{2-2}(\beta) \,
\ud \cos\beta.
\label{eq:appb12}
\end{equation}
In this case, $\csc^2(\beta/2) d^L_{2 -2}(\beta)$ is a polynomial in
$\cos\beta$ of degree $L-1$ and has a root at $\cos\beta=1$. It can thus
be expanded as a series in $d^n_{1 -1}$ with non-zero coefficients for
$n=1, \dots, L-1$, and so the integral in equation~(\ref{eq:appb12})
vanishes for $L \le l$. The
summation $S_3$ arises when computing band-power variances from
equations~(\ref{eq:5-58}) and (\ref{eq:5-59}); it occurs with
prefactors $|\cle_{LM}|^2$ and $|\clb_{LM}|^2$.
Since we are assuming the weight function is band-limited,
$L \la l_{\rmn{max}}$ in $S_3$, so for $l \gg l_{\rmn{max}}$ the
contribution from $S_3$ vanishes.
The final summation we require in Section~\ref{subsec:rot} is
\begin{equation}
S_4  \equiv \sum_{l'} (2l'+1) \left( \begin{array}{ccc} l & l' & L \\
                                                       -1 & 1 & 0
                                     \end{array} \right)
				\left( \begin{array}{ccc} l & l' & L \\
                                                       -1 & -1 & 2
                                     \end{array} \right),
\label{eq:appb13}
\end{equation}
which reduces to
\begin{equation}
S_4 = \frac{1}{2}\int\csc^2(\beta/2) d^{l}_{1 1}(\beta) d^L_{2 0}(\beta) \,
\ud \cos\beta.
\label{eq:appb14}
\end{equation}
This time, $\csc^2(\beta/2) d^L_{20}(\beta)$ is a polynomial in $\cos\beta$
of degree $L-1$ that has a root at $\cos\beta=-1$. The summation
$S_4$ can thus be expanded
in the $d^n_{11}$ with $n=1,\dots,L-1$, so that, like $S_3$,
it vanishes for $L\le l$. The summation arises from the cross term
between $|\eth w|^2_{LM}$ and $\cle_{LM}$ in
equations~(\ref{eq:5-58}) and (\ref{eq:5-59}), and
so this contribution to the band-power variance can also be assumed to vanish
for $l \gg l_{\rmn{max}}$.

\bsp  
\label{lastpage}

\begin{thebibliography}{}

\bibitem[\protect\citeauthoryear{Barkats et al.}{2005}]{barkats04}
	Barkats D., et al., 2005, ApJ, 619, L127

\bibitem[\protect\citeauthoryear{Bartolo et al.}{2004}]{bartolo04}
	Bartolo N., Komatsu E., Matarrese S., Rioto A., 2004,
	Phys.\ Rep., 402, 103

\bibitem[\protect\citeauthoryear{Bond, Jaffe \& Knox}
{Bond et al.}{1998}]{bond98}
	Bond J.\ R., Jaffe A.\ H., Knox L., 1998, Phys.\ Rev.\ D, 57,
	2117

\bibitem[\protect\citeauthoryear{Bond, Jaffe \& Knox}
{Bond et al.}{2000}]{bond00}
	Bond J.\ R., Jaffe A.\ H., Knox L., 2000, ApJ, 533, 19

\bibitem[\protect\citeauthoryear{Bunn et al.}{2003}]{bunn03}
	Bunn E.\ F., Zaldarriaga M., Tegmark M., de Oliveira-Costa A.,
	2003, Phys.\ Rev.\ D, 67, 3501

\bibitem[\protect\citeauthoryear{Chon}{2003}]{chon_thesis}
	Chon G., 2003, PhD thesis, University of Cambridge

\bibitem[\protect\citeauthoryear{Chon et al.}{2004}]{chon04}
	Chon G., Challinor A., Prunet S., Hivon E., Szapudi I., 2004,
	MNRAS, 350, 914

\bibitem[\protect\citeauthoryear{Crittenden et al.}{2002}]{crittenden02}
	Crittenden R.\ G., Natarajan P., Pen U.-L., Theuns T., 2002,
	ApJ, 568, 20

\bibitem[\protect\citeauthoryear{Efstathiou}{2004}]{efstathiou04}
	Efstathiou G., 2004, MNRAS, 349, 603

\bibitem[\protect\citeauthoryear{G\'{o}rski}{1994}]{gorski94}
	G\'{o}rski K.\ M., 1994, ApJ, 430, L85

\bibitem[\protect\citeauthoryear{G\'{o}rski et al.}{2004}]{gorski04}
	G\'{o}rski K.\ M., Hivon E., Banday A.\ J., Wandelt B.\ D.,
	Hansen F.\ K., Reinecke M., Bartelman M., 2004, preprint
	(astro-ph/0409513)

\bibitem[\protect\citeauthoryear{Hansen \& G\'{o}rski}{2003}]{hansen02b}
	Hansen F.\ K., G\'{o}rski K.\ M., 2003, MNRAS, 343, 559

\bibitem[\protect\citeauthoryear{Hansen, G\'{o}rski \& Hivon}
{Hansen et al.}{2002}]{hansen02}
	Hansen F.\ K., G\'{o}rski K.\ M., Hivon E., 2002, MNRAS, 336, 1304 

\bibitem[\protect\citeauthoryear{Hinshaw et al.}{2003}]{hinshaw03}
	Hinshaw G., et al., 2003, ApJS, 148, 135

\bibitem[\protect\citeauthoryear{Hivon et al.}{2002}]{hivon02}
	Hivon E., G\'{o}rski K.\ M., Netterfield C.\ B., Crill B.\ P.,
	Prunet S., Hansen F., 2002, ApJ, 567, 2

\bibitem[\protect\citeauthoryear{Jaffe, Kamionkowski \& Wang}{2000}]{jaffe00}
	Jaffe A.\ H., Kamionkowski M., Wang L., 2000, Phys.\ Rev.\ D,
	61, 083501

\bibitem[\protect\citeauthoryear{Kamionkowski, Kosowsky \& Stebbins}
{Kamionkowski et al.}{1997}]{kamionkowski97}
	Kamionkowski M., Kosowsky A., Stebbins A., 1997, Phys. Rev.
	Lett., 78, 2058

\bibitem[\protect\citeauthoryear{Kesden, Cooray \& Kamionkowski}
{Kesden et al.}{2002}]{kesden02}
	Kesden M., Cooray A., Kamionkowski M., 2002, Phys. Rev.
	Lett., 89, 011304

\bibitem[\protect\citeauthoryear{Knox \& Song}{2002}]{knox02}
	Knox L., Song Y.-S., 2002, Phys.\ Rev.\ Lett., 89, 1303

\bibitem[\protect\citeauthoryear{Kogut et al.}{2003}]{kogut03}
Kogut A. et al., 2003, ApJS, 148, 161

\bibitem[\protect\citeauthoryear{Kovac et al.}{2002}]{kovac02}
Kovac J., Leitch E.\ M., Pryke C., Carlstrom J.\ E., Halverson N.\ W.,
Holzapfel W.\ L., 2002, Nature, 420, 772

\bibitem[\protect\citeauthoryear{Leitch et al.}{2004}]{leitch04}
Leitch E.\ M., Kovac J.\ M., Halverson N.\ W., Carlstrom J.\ E.,
Pryke C., Smith M.\ W.\ E., 2004, preprint (astro-ph/0409357)

\bibitem[\protect\citeauthoryear{Lewis}{2003}]{lewis03}
	Lewis A., 2003, Phys.\ Rev.\ D, 68, 3509

\bibitem[\protect\citeauthoryear{Lewis, Challinor \& Lasenby}
{Lewis et al.}{2000}]{lewis00}
	Lewis A., Challinor A., Lasenby A., 2000, ApJ, 538, 473

\bibitem[\protect\citeauthoryear{Lewis, Challinor \& Turok}
{Lewis et al.}{2002}]{lewis02}
	Lewis A., Challinor A., Turok N., 2002, Phys.\ Rev.\ D, 65, 023505

\bibitem[\protect\citeauthoryear{Martin \& Schwarz}{2000}]{martin00}
	Martin J., Schwarz D., 2000, Phys.\ Rev.\ D, 62, 103520

\bibitem[\protect\citeauthoryear{Mollerach, Harari \& Matarrese}
{Mollerach et al.}{2004}]{mollerach04}
	Mollerach S., Harari D., Matarrese S., 2004, Phys.\ Rev.\ D,
	69, 063002

\bibitem[\protect\citeauthoryear{Netterfield et al.}{2002}]{netterfield02}
	Netterfield C.\ B., et al., 2002, ApJ, 571, 604

\bibitem[\protect\citeauthoryear{Newman \& Penrose}{1966}]{newman66}
	Newman E., Penrose R., 1966, J.\ Math.\ Phys., 863

\bibitem[\protect\citeauthoryear{Ponthieu et al.}{2005}]{ponthieu05}
	Ponthieu N., et al., 2005, preprint (astro-ph/0501427)

\bibitem[\protect\citeauthoryear{Readhead et al.}{2004}]{readhead04}
        Readhead A.\ C.\ S., et al., 2004, Science, 306, 836

\bibitem[\protect\citeauthoryear{Schulten \& Gordon}{1976}]{schulten76}
	Schulten K., Gordon R.\ G., 1976, Comput.\ Phys.\ Commun., 11, 269

\bibitem[\protect\citeauthoryear{Seljak \& Zaldarriaga}{1997}]{seljak97}
	Seljak U., Zaldarriaga M., 1997, Phys.\ Rev.\ Lett., 78, 2054

\bibitem[\protect\citeauthoryear{Seljak \& Hirata}{2004}]{seljak03}
	Seljak U., Hirata C.\ M., 2004, Phys.\ Rev.\ D, 69, 3005

\bibitem[\protect\citeauthoryear{Smith, Hu \& Kaplinghat}
{Smith et al.}{2004}]{smith04}
	Smith K.\ M., Hu W., Kaplinghat, M., 2004, Phys.\ Rev.\ D, 70, 043002

\bibitem[\protect\citeauthoryear{Spergel et al.}{2003}]{spergel03}
	Spergel D., et al., 2003, ApJS, 148, 175 

\bibitem[\protect\citeauthoryear{Szapudi, Prunet \& Colombi}
{Szapudi et al.}{2001}]{szapudi01}
	Szapudi I., Prunet S., Colombi S., 2001, ApJ, 561, L11

\bibitem[\protect\citeauthoryear{Tegmark \& de Oliveira-Costa}{2002}]
{tegmark02}
	Tegmark M., de Oliveira-Costa A., 2002, Phys.\ Rev.\ D, 64, 063001

\bibitem[\protect\citeauthoryear{Varshalovich, Moskalev \& Khersonskii}
{Varshalovich et al.}{1988}]{varshalovich}
	Varshalovich D.\ A., Moskalev A.\ N., Khersonskii V.\ K.,
	1988, Quantum Theory of Angular Momentum. World Scientific,
	Singapore

\bibitem[\protect\citeauthoryear{Wandelt, Hivon \& G\'{o}rski}
{Wandelt et al.}{2001}]{wandelt01b}
	Wandelt B.\ D., Hivon E., G\'{o}rski K.\ M., 2001,
	Phys.\ Rev.\ D, 64, 083003

\bibitem[\protect\citeauthoryear{Zaldarriaga \& Seljak}{1998}]{zaldarriaga98}
	Zaldarriaga M., Seljak U., 1998, Phys.\ Rev.\ D, 58, 023003

\end{thebibliography}
\end{document}